\documentclass
	[
		10 pt
	]
	{article}

\usepackage
	[
		a4paper,
		left = 2.500 cm,
		right = 2.500 cm,
		top = 2.500 cm,
		bottom = 3.000 cm
	]
	{geometry}
\usepackage{tikz, pgf, pgfplots}
\usetikzlibrary
	{
		3d,
		arrows,											
		arrows.meta,
		automata,										
		backgrounds,
		bending,
		calc,
		chains,
		datavisualization.formats.functions,			
		decorations.pathmorphing,
		decorations.pathreplacing,
		decorations.text,
		fadings,
		fit,
		graphs,
		graphs.standard,
		matrix,
		positioning,									
		quantikz2,
		quotes,
		shadings,
		shadows,
		shadows.blur,
		shapes,
		shapes.geometric,
		trees
	}
\pgfplotsset{compat = newest}
\usepgflibrary
	{
		shapes.misc
	}
\usepgfplotslibrary
	{
		dateplot
	}
\usepackage{tikzpeople}
\usepackage{fontawesome5}
\usepackage{smartdiagram}

\usepackage{amsthm}
\usepackage{amssymb}
\usepackage[bb = boondox]{mathalfa}
\usepackage{dsfont}
\usepackage{newtxmath}
\usepackage{mathtools}
\usepackage{multicol}
\usepackage{braket}
\usepackage{physics}
\numberwithin{equation}{section}
\usepackage{empheq}

\usepackage[ compat = newest ]{yquant}

\usepackage{sectsty}
\allsectionsfont{\color{WordBlue}}
\usepackage{xcolor}
\usepackage{graphicx}
\usepackage{varwidth}
\usepackage{wrapfig}
\usepackage{xurl}                                                                                  %
\usepackage{hyperref}
\hypersetup
	{
		colorlinks,
		linkcolor = {WordRed!75!black},
		citecolor = {WordBlueDarker25},
		urlcolor = {WordAquaDarker50}
	}
\usepackage{tabularray}
\usepackage{colortbl}
\usepackage{float}
\usepackage{nicematrix}
\usepackage{cellspace}
\usepackage{makecell}
\usepackage{multirow}
\usepackage{caption}
\usepackage[ linesnumbered, tworuled, vlined ] {algorithm2e}
\usepackage{appendix}
\usepackage{enumitem}
\usepackage{siunitx}
\usepackage{xintexpr}
\usepackage{spreadtab}
\usepackage{framed}
\usepackage{svg}
\usepackage{lipsum}
\definecolor{MyLightRed}{RGB}{244, 213, 245}
\definecolor{WordRed}{RGB}{255, 0, 102}
\definecolor{WordRedAccent5Lighter60}{HTML}{F5B5A7}
\definecolor{WordRedAccent5Darker25}{HTML}{B23214}
\definecolor{RedDarkLightest}{HTML}{ff0088}
\definecolor{RedDarkLight}{HTML}{ea005f}
\definecolor{RedPurple}{HTML}{aa007f}
\definecolor{Purple}{HTML}{911146}
\definecolor{PurpleDark}{RGB}{102, 0, 102}
\definecolor{WordLightGreen}{RGB}{140, 214, 192}
\definecolor{WordGreen}{RGB}{0, 176, 80}
\definecolor{GreenLightest}{HTML}{00ffa0}
\definecolor{GreenLighter1}{HTML}{00b383}
\definecolor{GreenLighter2}{HTML}{00aa7f}
\definecolor{GreenDark}{HTML}{225522}
\definecolor{GreenTeal}{HTML}{008080}
\definecolor{WordIceBlue}{RGB}{223, 227, 229}
\definecolor{MyVeryLightBlue}{RGB}{211, 245, 247}
\definecolor{WordBlueVeryLight}{RGB}{0, 176, 240}
\definecolor{WordBlueLight}{RGB}{0, 112, 192}
\definecolor{WordBlueDark}{RGB}{46, 116, 181}
\definecolor{WordBlueDarker}{RGB}{31, 78, 121}
\definecolor{WordBlueDarker25}{RGB}{54, 96, 146}
\definecolor{WordBlueDarker50}{RGB}{36, 64, 98}
\definecolor{WordBlueDarkest}{RGB}{0, 32, 96}
\definecolor{WordBlue}{RGB}{19, 65, 99}
\definecolor{MyBlue}{RGB}{0, 64, 128}
\definecolor{MyDarkBlue}{RGB}{0, 51, 102}
\definecolor{BlueVeryDark}{HTML}{222255}
\definecolor{MagentaVeryLight}{RGB}{178, 162, 201}
\definecolor{MagentaLighter}{RGB}{161, 106, 221}
\definecolor{MagentaLight}{RGB}{128, 100, 162}
\definecolor{MagentaDark}{RGB}{106, 65, 152}
\definecolor{MagentaVeryDark}{RGB}{97, 75, 128}
\definecolor{WordAquaLighter80}{RGB}{218, 238, 243}
\definecolor{WordAquaLighter60}{RGB}{183, 222, 232}
\definecolor{WordAquaLighter40}{RGB}{146, 205, 220}
\definecolor{WordAquaDarker25}{RGB}{49, 134, 155}
\definecolor{WordAquaAccent2Darker25}{HTML}{398E98}
\definecolor{WordAquaDarker50}{RGB}{33, 89, 103}
\definecolor{WordVeryLightTeal}{RGB}{223, 236, 235}
\definecolor{WordLightTeal}{RGB}{160, 199, 197}
\definecolor{WordDarkTealLighter80}{RGB}{207, 223, 234}
\definecolor{WordDarkTeal}{RGB}{72, 123, 119}
\definecolor{WordDarkerTeal}{RGB}{48, 82, 80}
\definecolor{WordTurquoiseLighter80}{RGB}{209, 238, 249}
\definecolor{WordGoldAccent1Lighter40}{HTML}{FFDF6A}
\definecolor{WordGoldAccent1Darker25}{HTML}{C49A00}
\definecolor{Brown}{HTML}{666633}
\definecolor{WordOrangeAccent2Lighter60}{HTML}{FCD3A4}
\definecolor{WordOrangeAccent4Lighter60}{HTML}{F7C5A1}
\definecolor{LavenderBlush}{RGB}{255, 240,  245}
\definecolor{MediumTurquoise}{RGB}{72, 209, 204}
\definecolor{PowderBlue}{RGB}{176, 224, 230}
\definecolor{SkyBlue}{RGB}{135, 206, 235}
\usepackage
	[
		skins,
		breakable,
		listings,
		raster,
		theorems
	]
	{tcolorbox}
\NewTcbTheorem
	[
		number within = section
	]
	{definition}
	{Definition}
	{
		breakable,
		skin = enhanced,
		enhanced jigsaw,			
		attach boxed title to top left = { xshift = -5.000 mm, yshift = 0.000 mm },
		boxed title style = { boxrule = 0.000 pt, sharp corners = all },
		varwidth boxed title,
		fonttitle = \bfseries,
		coltitle = SkyBlue!30!black,
		colbacktitle = SkyBlue!70!white,
		colframe = SkyBlue,
		colback = SkyBlue!20!white,
		sharp corners = all,
		toprule = 0.000 mm,
		bottomrule = 0.500 mm,
		leftrule = 0.500 mm,
		rightrule = 0.000 mm,
	}
	{def}
\NewTcbTheorem
	[
		number within = section
	]
	{example}
	{Example}
	{
		breakable,
		skin = enhanced,
		enhanced jigsaw,			
		attach boxed title to top left = { xshift = -5.000 mm, yshift = 0.000 mm },
		boxed title style = { boxrule = 0.000 pt, sharp corners = all },
		varwidth boxed title,
		fonttitle = \bfseries,
		coltitle = PowderBlue!30!black,
		colbacktitle = PowderBlue!70!white,
		colframe = PowderBlue,
		colback = white,
		sharp corners = all,
		toprule = 0.000 mm,
		bottomrule = 0.500 mm,
		leftrule = 0.500 mm,
		rightrule = 0.000 mm,
	}
	{xmp}

\newcounter{mathseed}
\pgfmathsetseed{\arabic{mathseed}}
\pgfdeclarelayer{background}
\pgfsetlayers{background, main}
\pgfdeclaredecoration{irregular fractal line}{init}
{
	\state{init}[width=\pgfdecoratedinputsegmentremainingdistance]
	{
		\pgfpathlineto{\pgfpoint{random*\pgfdecoratedinputsegmentremainingdistance}{(random*\pgfdecorationsegmentamplitude-0.02)*\pgfdecoratedinputsegmentremainingdistance}}
		\pgfpathlineto{\pgfpoint{\pgfdecoratedinputsegmentremainingdistance}{0pt}}
	}
}
\def\tornpaper#1{%
	\ifthenelse{\isodd{\value{mathseed}}}
	{%
		\tikz
		{
			\node[inner sep = 1em] (A) {#1};		
			\begin{pgfonlayer}{background}			
				\fill[paper]						
				\pgfextra{\pgfmathsetseed{\arabic{mathseed}}\addtocounter{mathseed}{1}}%
				{decorate[irregular cloudy border]{decorate{decorate{decorate{decorate[ragged border]{
										(A.north west) -- (A.north east)
				}}}}}}
				-- (A.south east)
				\pgfextra{\pgfmathsetseed{\arabic{mathseed}}}%
				{decorate[irregular spiky border]{decorate{decorate{decorate{decorate[ragged border]{
										-- (A.south west)
				}}}}}}
				-- (A.north west);
			\end{pgfonlayer}
		}
	}
	{%
		\tikz{
			\node[inner sep=1em] (A) {#1};  
			\begin{pgfonlayer}{background}  
				\fill[paper] 
				\pgfextra{\pgfmathsetseed{\arabic{mathseed}}\addtocounter{mathseed}{1}}%
				{decorate[irregular spiky border]{decorate{decorate{decorate{decorate[ragged border]{
										(A.north east) -- (A.north west)
				}}}}}}
				-- (A.south west)
				\pgfextra{\pgfmathsetseed{\arabic{mathseed}}}%
				{decorate[irregular cloudy border]{decorate{decorate{decorate{decorate[ragged border]{
										-- (A.south east)
				}}}}}}
				-- (A.north east);
		\end{pgfonlayer}}
	}
}
\title
	{
		A Quantum Algorithm for the Classification of Patterns of Boolean Functions
	}

\usepackage{orcidlink}
\newcommand{\orcidicon}[1]{\href{https://orcid.org/#1}{\includegraphics[height=\fontcharht\font`\B]{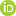}}}

\author
{
	Theodore Andronikos$^1$\orcidicon{0000-0002-3741-1271},
	Constantinos Bitsakos$^2$\orcidicon{0009-0003-3669-0453},
	Konstantinos Nikas$^2$\orcidicon{0000-0003-4424-9951},
	\\
	Georgios I. Goumas$^2$\orcidicon{0000-0001-7811-4831}
	and
	Nectarios Koziris$^2$\orcidicon{0000-0002-4890-8427}
	\\
	$^1$ \
	Department of Informatics, Ionian University, \\
	7 Tsirigoti Square, 49100 Corfu, Greece; \\
	andronikos@ionio.gr
	\\
	$^2$ \ 
	Computing Systems Laboratory, \\
	National Technical University of Athens, Greece; \\
	\{kbitsak, knikas, goumas, nkoziris\}@cslab.ece.ntua.gr
}
\begin{document}

\maketitle

\begin{abstract}
	This paper introduces a novel quantum algorithm that is able to classify a hierarchy of classes of imbalanced Boolean functions. The fundamental characteristic of imbalanced Boolean functions is that the proportion of elements in their domain that take the value $0$ is not equal to the proportion of elements that take the value $1$. For every positive integer $n$, the hierarchy contains a class of Boolean functions defined based on their behavioral pattern. The common trait of all the functions belonging to the same class is that they possess the same imbalance ratio. Our algorithm achieves classification in a straightforward manner as the final measurement reveals the unknown function with probability $1$. Let us also note that the proposed algorithm is an optimal oracular algorithm because it can classify the aforementioned functions with a single query to the oracle. At the same time we explain in detail the methodology we followed to design this algorithm in the hope that it will prove general and fruitful, given that it can be easily modified and extended to address other classes of imbalanced Boolean functions that exhibit different behavioral patterns.
\end{abstract}
\section{Introduction} \label{sec: Introduction}

The endeavor to construct quantum computers that surpass the capabilities of classical computers poses a significant challenge in our era. It is important to acknowledge that this goal has not yet been realized. However, substantial progress is evident, as illustrated by IBM's advancements with the 127-qubit Eagle \cite{IBMEagle2021}, the 433-qubit Osprey \cite{IBMOsprey2022}, the 1,121-qubit Condor \cite{IBMCondor2023}, and the latest and most powerful R2 Heron \cite{IBMHeron2024}. These developments suggest a swift movement towards the practical application of quantum technology. All these suggest that quantum technology has reached a level of maturity that warrants careful consideration in the development and implementation of algorithms targeting difficult problems.

The imperative to enhance the scale of quantum computers represents the most significant obstacle to their potential application in industrial-scale problems. It has become evident that advancing quantum computers beyond the Noisy Intermediate-Scale Quantum (NISQ) level will necessitate scientific breakthroughs and the resolution of various technological hurdles. In our assessment, the most promising strategy to address the scaling dilemma currently lies in the advancement of distributed quantum computing systems. In the realm of classical computing, the concept of interlinking smaller processors to distribute computational tasks emerged as a solution to scaling difficulties. This principle is believed to be equally relevant to quantum computing, where the scaling challenge encourages the exploration of connecting smaller quantum computers. A distributed quantum computing system would comprise a network of quantum nodes, each possessing a specific number of qubits for processing and the capability to transmit both classical and quantum information. Nevertheless, the inherent differences between quantum and classical computing introduce unique challenges, not present in classical networks, in the design of networked quantum computers. Fortunately, recently there have been significant technological advancements in hardware \cite{Photonic2024, NuQuantum2024} and design concepts \cite{Cacciapuoti2024, Illiano2024}. In fact, very recently researchers demonstrated distributed quantum computing by employing a photonic network interface to effectively connect two distinct quantum processors, thus, creating a unified and fully integrated quantum computer \cite{Main2025, OxfordNewsEvents2025}. It is our firm belief that we are entering the era of distributed quantum computing.

In this work, we introduce a new quantum algorithm that classifies classes of Boolean functions that are characterized by a specific patterns that demonstrate imbalance. The fundamental characteristic of these imbalanced Boolean functions is that the proportion of elements in their domain that take the value $0$ is not equal to the proportion of elements that take the value $1$. We refer to this algorithm as the Boolean Function Pattern Quantum Classifier, or BFPQC for short. We have drawn inspiration mainly from the many sophisticated works studying various extensions of the Deutsch-Jozsa algorithm. Already in \cite{Cleve1998}, the authors examined a multidimensional version of the Deutsch–Jozsa problem. This was further expanded in \cite{Chi2001} by considering evenly distributed and evenly balanced functions. Subsequently, in \cite{Holmes2003} the Deutsch–Jozsa algorithm was extended for balanced functions in finite Abelian subgroups. Another generalization appeared in \cite{Ballhysa2004}. Later, the researchers in \cite{Qiu2018} generalized the Deutsch–Jozsa problem and gave an optimal algorithm. A more recent clever generalization of the Deutsch–Jozsa algorithm can be found in \cite{OssorioCastillo2023}. Useful applications of the Deutsch–Jozsa algorithm were also obtained in \cite{Qiu2020} and in \cite{Zhengwei2021}. Two particularly interesting works towards establishing a distributed version of the Deutsch–Jozsa algorithm were \cite{Tanasescu2019} and \cite{Li2025}. In a related development, the authors in \cite{Nagata2020} extended Deutsch’s algorithm for binary Boolean functions. We should also mention that oracular algorithm geared towards computing Boolean functions or achieving classification are often encountered in the literature on Quantum Learning and Quantum Machine Learning. Some noteworthy studies in these areas include \cite{Bshouty1998, Farhi1999, Servedio2004, Hunziker2009, Yoo2014, Cross2015}. The fundamental characteristic of the Deutsch-Jozsa algorithm and its subsequent extensions is the distinction among constant and balanced functions, i.e., functions that the number of elements in their domain that take the value $0$ is  equal to the number of elements that take the value $1$. Here, to differentiate from this trend, we study imbalanced functions focusing on classifying specific patterns.

We present our algorithm in the form of game, featuring the familiar characters of Alice and Bob. It is anticipated that the entertaining aspect of games will facilitate a clearer understanding of the technical concepts presented. Since their introduction in 1999 \cite{Meyer1999,Eisert1999}, quantum games have gained considerable popularity, as quantum strategies often outperform classical ones \cite{Andronikos2018,Andronikos2021,Andronikos2022a}. A notable illustration of this is the well-known Prisoners' Dilemma \cite{Eisert1999}, which serves as a prime example and is applicable to various other abstract quantum games \cite{Giannakis2015a,Koh2024}. Furthermore, many classical systems can be transformed into quantum versions, including political frameworks, as demonstrated in recent studies \cite{Andronikos2022}. Especially cryptographic protocols like Quantum Key Distribution, Quantum Secret Sharing, Quantum Private Comparison, etc. are very often presented as interactions among signature players, including famous figures such as Alice, Bob, Charlie, Eve (see the recent works \cite{Ampatzis2021, Ampatzis2022, Ampatzis2023, Andronikos2023, Andronikos2023a, Andronikos2023b, Karananou2024, Andronikos2024, Andronikos2024a, Andronikos2024b, Andronikos2025}. In discussing games set within unconventional environments, it is noteworthy that games involving biological systems have garnered considerable interest \cite{Theocharopoulou2019,Kastampolidou2020a,Kostadimas2021}. It is particularly intriguing to note that biosystems can lead to biostrategies that may outperform traditional strategies, even in the renowned Prisoners' Dilemma game \cite{Kastampolidou2020,Kastampolidou2021,Papalitsas2021,Kastampolidou2023,Adam2023}.

\textbf{Contribution}.
Numerous sophisticated studies have been published in the literature that expand upon the Deutsch-Jozsa algorithm and explore balanced Boolean functions. However, as far as we are aware, there has been no previous research dedicated to imbalanced Boolean functions, which are characterized by an unequal number of elements in their domain that yield the values $0$ and $1$. This article introduces a novel quantum algorithm designed to classify a specific hierarchy of imbalanced Boolean function classes. For each positive integer $n$, this hierarchy includes a class of Boolean functions, which are defined according to their behavioral characteristics. A defining feature of all functions within the same class is their shared imbalance ratio. Our algorithm facilitates classification in a straightforward manner, as the final measurement determines the unknown function with a probability of $1$. It is important to highlight that the proposed algorithm is an optimal oracular algorithm, capable of classifying the specified functions with a single query to the oracle. Additionally, we provide a detailed explanation of the methodology employed in the development of this algorithm, with the expectation that it will prove both general and beneficial, as it can be readily adapted and expanded to tackle other classes of imbalanced Boolean functions that display varying behavioral patterns.

\subsection*{Organization} \label{subsec: Organization}

This article is structured in the following way. Section \ref{sec: Introduction} introduces the topic and includes references to relevant literature. Section \ref{sec: Notation & Terminology} offers a brief overview of key concepts, which serves as a basis for grasping the technical details. Section \ref{sec: The Basic Concepts Behind the BFPQC Algorithm} contains a comprehensive exposition to our algorithm including a detailed small scale example to build intuition. The general form of the algorithm is formally presented in Section \ref{sec: The General Form of the BFPQC Algorithm}. Finally, the paper wraps up with a summary and a discussion of the algorithm's nuances in Section \ref{sec: Discussion and Conclusions}.

\section{Notation \& terminology} \label{sec: Notation & Terminology}

\subsection{Boolean functions \& Oracles} \label{subsec: Boolean Functions & Oracles}

Let us first fix the notation and terminology we shall be using in the rest of this paper.

\begin{itemize}
	\item	
	$\mathbb{ B }$ is the binary set $\{ 0, 1 \}$.
	\item	
	A \emph{bit vector} $\mathbf{ b }$ of length $n$ is a sequence of $n$ bits: $\mathbf{ b } = b_{ n - 1 } \dots b_{ 0 }$. Two special bit vectors are the \emph{zero} and the \emph{one} bit vectors, denoted by $\mathbf{ 0 }$ and $\mathbf{ 1 }$, in which all the bits are zero and one, respectively: $\mathbf{ 0 } = 0 \dots 0$ and $\mathbf{ 1 } = 1 \dots 1$.
	\item	
	To make clear when we refer to a bit vector $\mathbf{ b } \in \mathbb{ B }^{ n }$, we write $\mathbf{ b }$ in boldface. Often, it is convenient to view $\mathbf{ b }$ as the binary representation of the integer $b$.
	\item	
	Each bit vector $\mathbf{ b } \in \mathbb{ B }^{ n }$ can also be viewed as a binary correspondence to one of the $2^{ n }$ basis kets that form the computational basis of the $2^{ n }$-dimensional Hilbert space.
\end{itemize}

\begin{definition} {Boolean Function} { Boolean Function}
	A \emph{Boolean} function $f$ is a function from $\mathbb{ B }^{ n }$ to $\mathbb{ B }$, $n \geq 1$.
\end{definition}

Oracles are an important concept in quantum computing and play a crucial role in many quantum algorithms. An oracle is a black box that encodes a specific function or information into a quantum circuit, allowing quantum algorithms to solve problems more efficiently than classical algorithms in certain cases. It is used to evaluate the function or check a condition without revealing the internal details of how the function works. In quantum algorithms, oracles are often used to mark solutions to a problem or to provide information about a function's behavior. For the purposes of our work, the following definition suffices.

\begin{definition} {Oracle \& Unitary Transform} { Oracle & Unitary Transform}
	An \emph{oracle} is a black box implementing a Boolean function $f$. The idea here is that, being a black box function, we know nothing about its inner working; just that it works correctly. Thus, it can be used for the construction of a unitary transform $U_{ f }$ that captures the behavior of $f$.

	Henceforth, we shall assume that the corresponding unitary transform $U_{ f }$ implements the standard schema
	\begin{align}
		\label{eq: Generic Unitary Transform U_f}
		U_{ f }
		\colon
		\ket{ y }
		\
		\ket{ \mathbf{ x } }
		\rightarrow
		\ket{ y \oplus f( \mathbf{ x } ) }
		\
		\ket{ \mathbf{ x } }
		\ .
	\end{align}
\end{definition}

In the literature this type of oracle is sometimes referred to as a Deutsch-Jozsa oracle. We note in passing that there also other variations of oracles, such as the Grover oracle, which is typically used to mark solutions to a problem. In this work, every oracle and unitary transform are assumed to satisfy \eqref{eq: Generic Unitary Transform U_f} and are used to deduce a function from its behavior. The standard measure of complexity in oracular algorithms is the query complexity, i.e., the number of queries to the oracle used by the algorithm.

For completeness, we recall the states $\ket{ + }$ and $\ket{ - }$, which are defined as

\begin{tcolorbox}
	[
		enhanced,
		breakable,
		center title,
		fonttitle = \bfseries,
		grow to left by = 0.000 cm,
		grow to right by = 0.000 cm,
		colback = white,			
		enhanced jigsaw,			
		sharp corners,
		toprule = 0.001 pt,
		bottomrule = 0.001 pt,
		leftrule = 0.001 pt,
		rightrule = 0.001 pt,
	]
	\begin{minipage} [ b ] { 0.450 \textwidth }
		\begin{align}
			\label{eq: Ket +}
			\ket{ + }
			=
			H
			\ket{ 0 }
			=
			\frac
			{ \ket{ 0 } + \ket{ 1 } }
			{ \sqrt{ 2 } }
		\end{align}
	\end{minipage}
	\begin{minipage} [ b ] { 0.450 \textwidth }
		\begin{align}
			\label{eq: Ket -}
			\ket{ - }
			=
			H
			\ket{ 1 }
			=
			\frac
			{ \ket{ 0 } - \ket{ 1 } }
			{ \sqrt{ 2 } }
		\end{align}
	\end{minipage}
\end{tcolorbox}

To obtain any useful information from the schema \eqref{eq: Generic Unitary Transform U_f}, we set $\ket{ y }$ equal to $\ket{ - }$, in which case \eqref{eq: Generic Unitary Transform U_f} takes the following familiar form:

\begin{align}
	\label{eq: Unitary Transform U_f}
	U_{ f }
	\colon
	\ket{ - }
	\
	\ket{ \mathbf{ x } }
	\rightarrow
	( - 1 )^{ f ( \mathbf{ x } ) }
	\
	\ket{ - }
	\
	\ket{ \mathbf{ x } }
	\ .
\end{align}

Figures \ref{fig: The Quantum Circuit for the Generic Unitary Transform U_f} and \ref{fig: The Quantum Circuit for the Unitary Transform U_f} give a visual outline of the unitary transforms $U_{ f }$ that implement schemata \eqref{eq: Generic Unitary Transform U_f} and \eqref{eq: Unitary Transform U_f}, respectively.

\begin{tcolorbox}
	[
		enhanced,
		breakable,
		center title,
		fonttitle = \bfseries,
		grow to left by = 0.000 cm,
		grow to right by = 0.000 cm,
		colback = MagentaLighter!03,
		enhanced jigsaw,			
		sharp corners,
		toprule = 0.001 pt,
		bottomrule = 0.001 pt,
		leftrule = 0.001 pt,
		rightrule = 0.001 pt,
	]
	\begin{minipage} [ b ] { 0.450 \textwidth }
		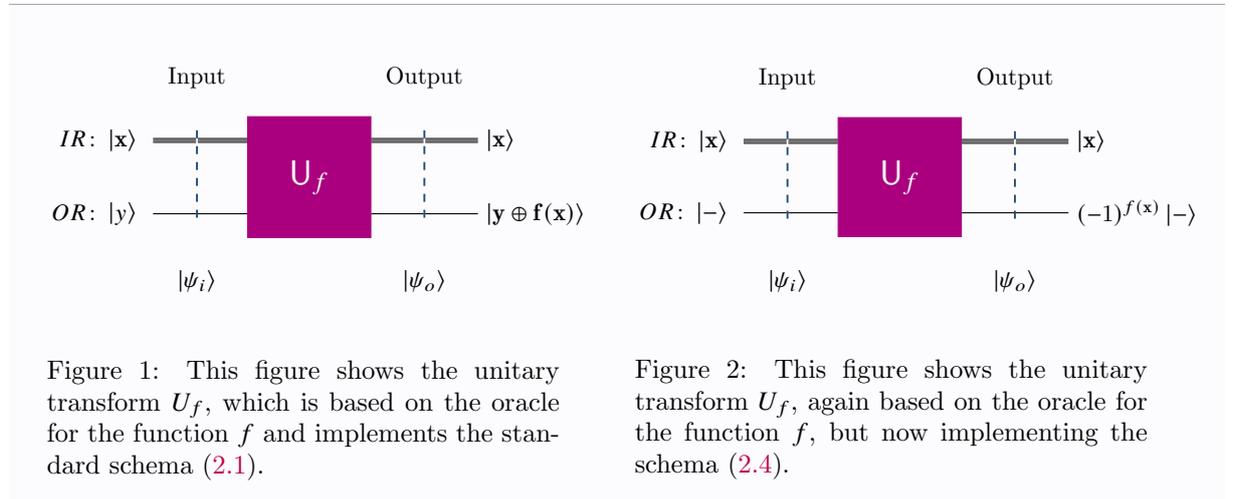
\begin{figure}[H]
			\centering
			\begin{tikzpicture} [ scale = 0.900 ] 
				\begin{yquant}
					qubits { $IR \colon \ket{ \mathbf{ x } }$ } IR;
					qubit { $OR \colon \ket{ y }$ } OR;
					[ name = Input, WordBlueDarker, line width = 0.250 mm, label = { [ label distance = 0.600 cm ] north: Input } ]
					barrier ( - ) ;
					hspace { 0.100 cm } IR;
					[ draw = RedPurple, fill = RedPurple, x radius = 0.900 cm, y radius = 0.700 cm ] box { \color{white} \Large \sf{U}$_{ f }$} (-);
					[ name = Output, WordBlueDarker, line width = 0.250 mm, label = { [ label distance = 0.600 cm ] north: Output } ]
					barrier ( - ) ;
					output { $\ket{ \mathbf{ x } }$ } IR;
					output { $\ket{ \mathbf{ y \oplus f ( \mathbf{ x } ) } }$ } OR;
					\node [ below = 1.250 cm ] at (Input) { $\ket{ \psi_{ i } }$ };
					\node [ below = 1.250 cm ] at (Output) { $\ket{ \psi_{ o } }$ };
				\end{yquant}
				\node [ anchor = center, below = 1.250 cm of Input ] (PhantomNode) { };
			\end{tikzpicture}
			\caption{This figure shows the unitary transform $U_{ f }$, which is based on the oracle for the function $f$ and implements the standard schema \eqref{eq: Generic Unitary Transform U_f}.}
			\label{fig: The Quantum Circuit for the Generic Unitary Transform U_f}
		\end{figure}
	\end{minipage}
	\hspace{ 0.750 cm }
	\begin{minipage} [ b ] { 0.450 \textwidth }
		\begin{figure}[H]
			\centering
			\begin{tikzpicture} [ scale = 0.900 ] 
				\begin{yquant}
					qubits { $IR \colon \ket{ \mathbf{ x } }$ } IR;
					qubit { $OR \colon \ket{ - }$ } OR;
					[ name = Input, WordBlueDarker, line width = 0.250 mm, label = { [ label distance = 0.600 cm ] north: Input } ]
					barrier ( - ) ;
					hspace { 0.100 cm } IR;
					[ draw = RedPurple, fill = RedPurple, x radius = 0.900 cm, y radius = 0.700 cm ] box { \color{white} \Large \sf{U}$_{ f }$} (-);
					[ name = Output, WordBlueDarker, line width = 0.250 mm, label = { [ label distance = 0.600 cm ] north: Output } ]
					barrier ( - ) ;
					output { $\ket{ \mathbf{ x } }$ } IR;
					output { $( - 1 )^{ f ( \mathbf{ x } ) } \ket{ - }$ } OR;
					\node [ below = 1.250 cm ] at (Input) { $\ket{ \psi_{ i } }$ };
					\node [ below = 1.250 cm ] at (Output) { $\ket{ \psi_{ o } }$ };
				\end{yquant}
				\node [ anchor = center, below = 1.250 cm of Input ] (PhantomNode) { };
			\end{tikzpicture}
			\caption{This figure shows the unitary transform $U_{ f }$, again based on the oracle for the function $f$, but now implementing the schema \eqref{eq: Unitary Transform U_f}.}
			\label{fig: The Quantum Circuit for the Unitary Transform U_f}
		\end{figure}
	\end{minipage}
\end{tcolorbox}

In all the quantum circuits used in this work, including those depicted in Figures \ref{fig: The Quantum Circuit for the Generic Unitary Transform U_f} and \ref{fig: The Quantum Circuit for the Unitary Transform U_f}, the following conventions are used.

\begin{itemize}
	\item	
	The way of ordering the qubits adheres to the Qiskit \cite{Qiskit2025}, i.e., the little-endian qubit indexing convention, where the least significant qubit is at the top of the figure and the most significant at the bottom.
	\item	
	$IR$ is the quantum input register that contains $n$ qubits.
	\item	
	$OR$ is the single-qubit output register that is initialized to an arbitrary state $\ket{ y }$ in Figure \ref{fig: The Quantum Circuit for the Generic Unitary Transform U_f} and to state $\ket{ - }$ in Figure \ref{fig: The Quantum Circuit for the Unitary Transform U_f}.
	\item	
	$U_{ f }$ is the unitary transform. Its precise mathematical expression depends on $f$ and is hidden. However, it is taken for granted that is satisfies relation \eqref{eq: Generic Unitary Transform U_f} in Figure \ref{fig: The Quantum Circuit for the Generic Unitary Transform U_f} and relation \eqref{eq: Unitary Transform U_f} in Figure \ref{fig: The Quantum Circuit for the Unitary Transform U_f}.
\end{itemize}

We mention that in the literature it is very common to use the word ``promise'' when we refer to a particular property of the Boolean function $f$, meaning that we are guaranteed, or, if you prefer we are certain with probability $1.0$, that $f$ satisfies the property in question. A prominent such example comes from the Deutsch–Jozsa algorithm, where we are given the promise that $f$ is either \emph{constant}, or \emph{balanced}.

Extending the operation of addition modulo $2$ to bit vectors is a natural and fruitful generalization.

\begin{definition} {Bitwise Addition Modulo $2$} { Bitwise Addition Modulo $2$}
	Given two bit vectors $\mathbf{ x }, \mathbf{ y } \in \mathbb{ B }^{ n }$, with $\mathbf{ x } = x_{ n - 1 } \dots x_{ 0 }$ and $\mathbf{ y } = y_{ n - 1 } \dots y_{ 0 }$, we define their \emph{bitwise sum modulo} $2$, denoted by $\mathbf{ x } \oplus \mathbf{ y }$, as
	\begin{align}
		\label{eq: Bitwise Addition Modulo $2$}
		\mathbf{ x }
		\oplus
		\mathbf{ y }
		\coloneq
		( x_{ n - 1 } \oplus y_{ n - 1 } )
		\dots
		( x_{ 0 } \oplus y_{ 0 } )
		\ .
	\end{align}
\end{definition}

Following the standard approach, we use the same symbol $\oplus$ to denote the operation of addition modulo $2$ two between bits, and the bitwise sum modulo $2$ between two bit vectors because the context always makes clear the intended operation.

\section{The basic concepts behind the BFPQC algorithm} \label{sec: The Basic Concepts Behind the BFPQC Algorithm}

In this paper we introduce a new quantum algorithm that differentiates and classifies a class of Boolean function that are characterized by a specific collection of patterns demonstrating imbalance. In view of its intended purpose, we call this algorithm the Boolean Function Pattern Quantum Classifier, or BFPQC for short. The current section gives the definitions regarding the main concepts, and presents a toy scale example illustrating its operation.

\begin{tcolorbox}
	[
		enhanced,
		breakable,
		center title,
		fonttitle = \bfseries,
		colbacktitle = azure4,
		coltitle = white,
		title = The purpose of the classification algorithm,
		grow to left by = 0.000 cm,
		grow to right by = 0.000 cm,
		colframe = azure1,
		colback = azure9!75,
		enhanced jigsaw,			
		sharp corners,
		boxrule = 0.500 pt,
	]
	In this paper we introduce an exact quantum algorithm that classifies a hierarchy of classes of Boolean functions. The algorithm can distinguish any two Boolean functions in this hierarchy, provided one is not the negation of the other, by giving rise to different elements of the computational basis with probability $1$. As expected, the algorithm can't distinguish a function from its negation, as they are both associated to the same basis ket. Our algorithm is an oracular algorithm because it relies on a oracle to achieve the classification. Its efficiency is demonstrated by the fact that it is optimal because it requires just one single query to complete its task.
\end{tcolorbox}

Here we solve what is commonly referred to in the quantum literature as a \emph{promise} problem, i.e., a problem where the input is promised to belong to a specific set. Promise algorithms are not required to work correctly on any input that doesn't satisfy the promise. Many quantum algorithms are designed to solve promise problems. For example, in the Deutsch-Jozsa algorithm, the promise is that the function is either constant or balanced. The algorithm is designed to distinguish between these two cases efficiently, but it doesn't need to handle functions that are neither constant nor balanced. The same applies to our case: the BFPQC algorithm can correctly handle any function that belongs to a rigorously defined hierarchy, but it will not output the correct answer if this is not the case.

Our plan of action consists of the following successive steps.

\begin{enumerate}
	[ left = 0.600 cm, labelsep = 0.750 cm, start = 1 ]
	\renewcommand \labelenumi { (\textbf{S}$_{ \theenumi }$) }
	\item	We focus on imbalanced Boolean functions, i.e., those with the property that the number of elements in their domain that take the value $0$ is not equal to the number of elements that take the value $1$.
	\item	We employ the concept of pattern vectors to capture the behavior of imbalanced Boolean functions. For each positive integer $n \geq 1$ we define a set of $2^{ 2 n }$ pattern vectors that all have equal imbalance ratio, which is always $< \frac { 1 } { 2 }$.
	\item	Identifying an appropriate set of pattern vector enables the construction of the corresponding unitary transform that accomplishes the classification.
\end{enumerate}

\begin{definition} {Pattern Vector} { Pattern Vector}
	Given the Boolean function $f \colon \mathbb{ B }^{ n } \rightarrow \mathbb{ B }$, $n \geq 1$, we define the concept of the unique \emph{pattern vector} that encodes the behavior of $f$.
	\begin{itemize}
		\item	
		The pattern vector $\mathbf{ p }$ $=$ $p_{ 2^{ n } - 1 }$ $\dots$ $p_{ 1 }$ $p_{ 0 }$ of $f$ is the element of $\mathbb{ B }^{ 2^{ n } }$, such that $p_{ i } = f ( \mathbf{ i } )$, where $\mathbf{ i }$ is the binary bit vector representing integer $i$, $0 \leq i \leq 2^{ n } - 1$. In other words, the pattern vector $\mathbf{ p }$ lists the binary values of $f ( \mathbf{ i } )$ as $\mathbf{ i }$ ranges over $\mathbb{ B }^{ n }$. To enhance comprehension, we visualize the details below.

				\begin{tblr}
					{
						colspec =
						{
							Q [ r, m, 4.000 cm ]
							Q [ c, m, 1.400 cm ]
							Q [ c, m, 0.350 cm ]
							Q [ c, m, 1.400 cm ]
							Q [ c, m, 0.000 cm ]
							Q [ c, m, 1.400 cm ]
						},
						rowspec =
						{
							| [ 6.000 pt, SkyBlue!20!white ]
							Q
							Q
							Q
							Q
							Q
							Q
							Q
							| [ 6.000 pt, SkyBlue!20!white ]
						}
					}
					\emph{Position:}
					&
					\SetCell { bg = purple7, fg = black } $2^{ n } - 1$
					&
					\dots
					&
					\SetCell { bg = purple8, fg = black } $1$
					&
					&
					\SetCell { bg = purple9, fg = black } $0$
					\\
					&
					$\downarrow$
					&
					\dots
					&
					$\downarrow$
					&
					&
					$\downarrow$
					\\
					\emph{Position in binary:}
					&
					\SetCell { bg = teal7, fg = black } $\mathbf{ 1 \dots 1 1 }$
					&
					\dots
					&
					\SetCell { bg = teal8, fg = black } $\mathbf{ 0 \dots 0 1 }$
					&
					&
					\SetCell { bg = teal9, fg = black } $\mathbf{ 0 \dots 0 0 }$
					\\
					&
					$\downarrow$
					&
					\dots
					&
					$\downarrow$
					&
					&
					$\downarrow$
					\\
					\emph{Function value:}
					&
					\SetCell { bg = cyan7, fg = black } $f ( \mathbf{ 1 \dots 1 1 } )$
					&
					\dots
					&
					\SetCell { bg = cyan8, fg = black } $f ( \mathbf{ 0 \dots 0 1 } )$
					&
					&
					\SetCell { bg = cyan9, fg = black } $f ( \mathbf{ 0 \dots 0 0 } )$
					\\
					&
					$\downarrow$
					&
					\dots
					&
					$\downarrow$
					&
					&
					$\downarrow$
					\\
					\emph{Pattern bit:}
					&
					\SetCell { bg = violet7, fg = black } $p_{ 2^{ n } - 1 }$
					&
					\dots
					&
					\SetCell { bg = violet8, fg = black } $p_{ 1 }$
					&
					&
					\SetCell { bg = violet9, fg = black } $p_{ 0 }$
				\end{tblr}
		\item	
		Given the pattern vector $\mathbf{ p }$ $=$ $p_{ 2^{ n } - 1 }$ $\dots$ $p_{ 1 }$ $p_{ 0 }$ of $f$, its \emph{negation}, denoted by $\overline { \mathbf{ p } }$, is the pattern vector $\overline { p_{ 2^{ n } - 1 } }$ $\dots$ $\overline { p_{ 1 } }$ $\overline { p_{ 0 } }$, which corresponds to the function $\overline { f }$.
	\end{itemize}
\end{definition}

It is clear by the preceding Definition \ref{def: Pattern Vector} that there is a one to one correspondence between Boolean functions and patterns vectors. We could say that a Boolean function and its pattern vector are the two sides of the same coin. Therefore, just as knowing the behavior of a Boolean function enables the construction of its pattern vector, conversely, the pattern vector contains all the information necessary to reconstruct the Boolean function. This duality is emphasized by the next Figure \ref{fig: Boolean Function - Pattern Vector Equivalence}.

\begin{tcolorbox}
	[
		enhanced,
		breakable,
		center title,
		fonttitle = \bfseries,
		grow to left by = 0.000 cm,
		grow to right by = 0.000 cm,
		colback = white,			
		enhanced jigsaw,			
		sharp corners,
		toprule = 0.001 pt,
		bottomrule = 0.001 pt,
		leftrule = 0.001 pt,
		rightrule = 0.001 pt,
	]
	\begin{figure}[H]
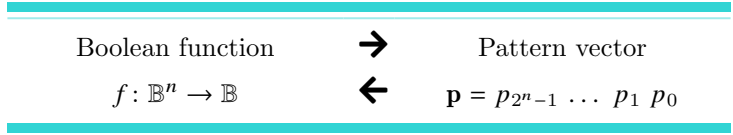

		\centering
		\begin{tblr}
			{
				colspec =
				{
					Q [ c, m, 4.000 cm ]
					Q [ c, m, 0.250 cm ]
					Q [ c, m, 4.000 cm ]
				},
				rowspec =
				{
					| [ 3.500 pt, cyan8 ]
					| [ 0.750 pt, cyan9 ]
					| [ 0.350 pt, white ]
					Q
					Q
					| [ 0.350 pt, white ]
					| [ 3.500 pt, cyan8 ]
				}
			}
			Boolean function
			&
			{ \large \faArrowRight }
			&
			Pattern vector
			\\
			$f \colon \mathbb{ B }^{ n } \rightarrow \mathbb{ B }$
			&
			{ \large \faArrowLeft }
			&
			$\mathbf{ p }$ $=$ $p_{ 2^{ n } - 1 }$ $\dots$ $p_{ 1 }$ $p_{ 0 }$
		\end{tblr}
		\caption{The duality between Boolean functions and their pattern vectors.}
		\label{fig: Boolean Function - Pattern Vector Equivalence}
	\end{figure}
\end{tcolorbox}

\begin{definition} {Equivalent \& Orthogonal Pattern Vectors} { Equivalent & Orthogonal Pattern Vectors}
	Consider the distinct pattern vectors $\mathbf{ p }$ and $\mathbf{ q }$, corresponding to the Boolean functions $f, g \colon \mathbb{ B }^{ n } \rightarrow \mathbb{ B }$.
	\begin{itemize}
		\item	
		$\mathbf{ p }$ and $\mathbf{ q }$ are \emph{equivalent} if they satisfy the following relation:
				\begin{align}
					\label{eq: Equivalent & Orthogonal Pattern Vectors}
					\mathbf{ p }
					\oplus
					\mathbf{ q }
					=
					\mathbf{ 1 }
					\ .
				\end{align}
		\item	
		$\mathbf{ p }$ and $\mathbf{ q }$ are \emph{orthogonal} if $\mathbf{ p } \oplus \mathbf{ q }$ contains $2^{ n - 1 }$ $0$s and $2^{ n - 1 }$ $1$s.
	\end{itemize}
\end{definition}

\begin{definition} {Imbalance Ratio} { Imbalance Ratio}
	Given a pattern vector $\mathbf{ p }$ of length $2^{ n }$, let $\vmathbb{ 0 }_{ \mathbf{ p } }$ and $\mathds{ 1 }_{ \mathbf{ p } }$ denote the number of $0$s and $1$s appearing in $\mathbf{ p }$. The \emph{imbalance ratio} of $\mathbf{ p }$ is defined as
	\begin{align}
		\label{eq: Imbalance Ratio}
		\rho
		\coloneq
		\min
		\left\{
		\frac { \vmathbb{ 0 }_{ \mathbf{ p } } } { 2^{ n } },
		\frac { \mathds{ 1 }_{ \mathbf{ p } } } { 2^{ n } }
		\right\}
		\ .
	\end{align}
	If $\mathbf{ p }$ is the pattern vector of $f$, we shall also say that $\rho$ is the imbalance ratio of $f$. In the same spirit, if $P$ and $F$ are a collection of pattern vectors and a collection of Boolean functions with common imbalance ratio $\rho$, respectively, we will speak of $\rho$ being the imbalance ratio of $P$ and $F$.
\end{definition}

As we pointed out previously, we visualize the execution of the BFPQC algorithm as the evolution of a game played between our prolific stars Alice and Bob, according to the following rules.

\begin{enumerate}
	[ left = 0.600 cm, labelsep = 0.750 cm, start = 1 ]
	\renewcommand \labelenumi { (\textbf{G}$_{ \theenumi }$) }
	\item	Bob is free to choose any Boolean function, provided that it belongs to the promised class of functions.
	\item	Bob wins the game if Alice fails to recognize the chosen function with one try. Otherwise, Alice is the winner.
	\item	In terms of implementing the game as a quantum circuit, Bob chooses the hidden oracle, while Alice furnishes the classifier.
\end{enumerate}

Before we proceed to introduce more technical machinery, we give a toy scale example to build intuition.

\begin{example} {A Toy Scale Example} { A Toy Scale Example - 1}
	Let us consider the following two families of Boolean functions defined on $\mathbb{ B }^{ 2 }$.

\begin{tcolorbox}
	[
		enhanced,
		breakable,
		center title,
		fonttitle = \bfseries,
		grow to left by = 0.000 cm,
		grow to right by = 0.000 cm,
		colback = white,			
		enhanced jigsaw,			
		sharp corners,
		boxrule = 0.001 pt,
	]
	\begin{minipage} [ b ] { 0.450 \textwidth }
		\begin{align}
			\label{eq: 2D 1N-3P Boolean Functions}
			\left
			\{
			\
			\begin{aligned}
				f_{ 0 }
				(
				x_{ 1 },
				x_{ 0 }
				)
				&\coloneq
				\overline{ x_{ 1 } }
				\wedge
				\overline{ x_{ 0 } }
				\\
				f_{ 1 }
				(
				x_{ 1 },
				x_{ 0 }
				)
				&\coloneq
				\overline{ x_{ 1 } }
				\wedge
				x_{ 0 }
				\\
				f_{ 2 }
				(
				x_{ 1 },
				x_{ 0 }
				)
				&\coloneq
				x_{ 1 }
				\wedge
				\overline{ x_{ 0 } }
				\\
				f_{ 3 }
				(
				x_{ 1 },
				x_{ 0 }
				)
				&\coloneq
				x_{ 1 }
				\wedge
				x_{ 0 }
				\\
			\end{aligned}
			\
			\right
			\}
		\end{align}
	\end{minipage}
	\hfill			
	\begin{minipage} [ b ] { 0.450 \textwidth }
		\begin{align}
			\label{eq: 2D 1P-3N Boolean Functions}
			\left
			\{
			\
			\begin{aligned}
				g_{ 0 }
				(
				x_{ 1 },
				x_{ 0 }
				)
				&\coloneq
				x_{ 1 }
				\vee
				x_{ 0 }
				\\
				g_{ 1 }
				(
				x_{ 1 },
				x_{ 0 }
				)
				&\coloneq
				x_{ 1 }
				\vee
				\overline{ x_{ 0 } }
				\\
				g_{ 2 }
				(
				x_{ 1 },
				x_{ 0 }
				)
				&\coloneq
				\overline{ x_{ 1 } }
				\vee
				x_{ 0 }
				\\
				g_{ 3 }
				(
				x_{ 1 },
				x_{ 0 }
				)
				&\coloneq
				\overline{ x_{ 1 } }
				\vee
				\overline{ x_{ 0 } }
				\\
			\end{aligned}
			\
			\right
			\}
		\end{align}
	\end{minipage}
\end{tcolorbox}

Their truth values and pattern vectors are given in Tables \ref{tbl: Truth Values & Pattern Vectors of 2D 1N-3P Boolean Functions} and \ref{tbl: Truth Values & Pattern Vectors of 2D 1P-3N Boolean Functions} below.

\begin{tcolorbox}
	[
		enhanced,
		breakable,
		center title,
		fonttitle = \bfseries,
		grow to left by = 0.500 cm,
		grow to right by = 0.500 cm,
		colback = white,			
		enhanced jigsaw,			
		sharp corners,
		toprule = 0.001 pt,
		bottomrule = 0.001 pt,
		leftrule = 0.001 pt,
		rightrule = 0.001 pt,
	]
	\begin{table}[H]
		\begin{minipage} [ t ] { 0.485 \textwidth }
			\caption{The truth values and the pattern vectors of $f_{ 0 }, f_{ 1 }, f_{ 2 }$, and $f_{ 3 }$.}
			\label{tbl: Truth Values & Pattern Vectors of 2D 1N-3P Boolean Functions}
			\centering
			\SetTblrInner { rowsep = 1.000 mm }
			\begin{tblr}
				{
					colspec =
					{
						Q [ c, m, 1.000 cm ]
						| [ 1.000 pt, purple7 ]
						| [ 1.000 pt, purple7 ]
						Q [ c, m, 0.500 cm ]
						| [ 0.500 pt, purple7 ]
						Q [ c, m, 0.500 cm ]
						| [ 0.500 pt, purple7 ]
						Q [ c, m, 0.500 cm ]
						| [ 0.500 pt, purple7 ]
						Q [ c, m, 0.500 cm ]
						| [ 1.000 pt, purple7 ]
						Q [ c, m, 1.500 cm ]
					},
					rowspec =
					{
						|
						[ 3.500 pt, purple7 ]
						|
						[ 0.750 pt, purple7 ]
						|
						[ 0.250 pt, white ]
						Q
						|
						[ 0.500 pt, purple7 ]
						Q
						|
						[ 0.500 pt, purple7 ]
						Q
						|
						[ 0.500 pt, purple7 ]
						Q
						|
						[ 0.500 pt, purple7 ]
						Q
						|
						[ 3.500 pt, purple7 ]
					}
				}
				&
				\SetCell { bg = purple5, fg = white } $\mathbf{ 00 }$
				&
				\SetCell { bg = purple5, fg = white } $\mathbf{ 01 }$
				&
				\SetCell { bg = purple5, fg = white } $\mathbf{ 10 }$
				&
				\SetCell { bg = purple5, fg = white } $\mathbf{ 11 }$
				&
				\SetCell { bg = purple5, fg = white, font = \bfseries } {Pattern\\Vector}
				\\
				$f_{ 0 }$
				&
				$1$
				&
				$0$
				&
				$0$
				&
				$0$
				&
				$0001$
				\\
				$f_{ 1 }$
				&
				$0$
				&
				$1$
				&
				$0$
				&
				$0$
				&
				$0010$
				\\
				$f_{ 2 }$
				&
				$0$
				&
				$0$
				&
				$1$
				&
				$0$
				&
				$0100$
				\\
				$f_{ 3 }$
				&
				$0$
				&
				$0$
				&
				$0$
				&
				$1$
				&
				$1000$
				\\
			\end{tblr}
		\end{minipage}
		\hspace{ 0.500 cm }
		\begin{minipage} [ t ] { 0.485 \textwidth }
			\caption{The truth values and the pattern vectors of  $g_{ 0 }, g_{ 1 }, g_{ 2 }$, and $g_{ 3 }$.}
			\label{tbl: Truth Values & Pattern Vectors of 2D 1P-3N Boolean Functions}
			\centering
			\SetTblrInner { rowsep = 1.000 mm }
			\begin{tblr}
				{
					colspec =
					{
						Q [ c, m, 1.000 cm ]
						| [ 1.000 pt, teal7 ]
						| [ 1.000 pt, teal7 ]
						Q [ c, m, 0.500 cm ]
						| [ 0.500 pt, teal7 ]
						Q [ c, m, 0.500 cm ]
						| [ 0.500 pt, teal7 ]
						Q [ c, m, 0.500 cm ]
						| [ 0.500 pt, teal7 ]
						Q [ c, m, 0.500 cm ]
						| [ 1.000 pt, teal7 ]
						Q [ c, m, 1.500 cm ]
					},
					rowspec =
					{
						|
						[ 3.500 pt, teal7 ]
						|
						[ 0.750 pt, teal7 ]
						|
						[ 0.250 pt, white ]
						Q
						|
						[ 0.500 pt, teal7 ]
						Q
						|
						[ 0.500 pt, teal7 ]
						Q
						|
						[ 0.500 pt, teal7 ]
						Q
						|
						[ 0.500 pt, teal7 ]
						Q
						|
						[ 3.500 pt, teal7 ]
					}
				}
				&
				\SetCell { bg = GreenLighter2, fg = white } $\mathbf{ 00 }$
				&
				\SetCell { bg = GreenLighter2, fg = white } $\mathbf{ 01 }$
				&
				\SetCell { bg = GreenLighter2, fg = white } $\mathbf{ 10 }$
				&
				\SetCell { bg = GreenLighter2, fg = white } $\mathbf{ 11 }$
				&
				\SetCell { bg = GreenLighter2, fg = white, font = \bfseries} {Pattern\\Vector}
				\\
				$g_{ 0 }$
				&
				$0$
				&
				$1$
				&
				$1$
				&
				$1$
				&
				$1110$
				\\
				$g_{ 1 }$
				&
				$1$
				&
				$0$
				&
				$1$
				&
				$1$
				&
				$1101$
				\\
				$g_{ 2 }$
				&
				$1$
				&
				$1$
				&
				$0$
				&
				$1$
				&
				$1011$
				\\
				$g_{ 3 }$
				&
				$1$
				&
				$1$
				&
				$1$
				&
				$0$
				&
				$0111$
				\\
			\end{tblr}
		\end{minipage}
	\end{table}
\end{tcolorbox}

The four functions $f_{ 0 }, f_{ 1 }, f_{ 2 }$, and $f_{ 3 }$ exhibit a common pattern, namely for precisely one element $\mathbf{ x } \in \mathbb{ B }^{ 2 }$ their value is $1$, while for the remaining three elements their value is $0$. Symmetrically, the four $g_{ 0 }, g_{ 1 }, g_{ 2 }$, and $g_{ 3 }$ functions exhibit an analogous motif, i.e., for precisely one element $\mathbf{ x } \in \mathbb{ B }^{ 2 }$ their value is $0$, while for the remaining three elements their value is $1$. Obviously, this is because $g_{ i } = \overline{ f_{ i } }$, $0 \leq i \leq 3$. The imbalance ratio $\rho$ for both families is the same, namely $\rho = \frac { 1 } { 4 }$. The four pattern vectors shown in Table \ref{tbl: Truth Values & Pattern Vectors of 2D 1N-3P Boolean Functions} are pairwise orthogonal and constitute the set $P_{ 2 } = \{ 1000, 0100, 0010, 0001 \}$. The same holds for the four pattern vectors in Table \ref{tbl: Truth Values & Pattern Vectors of 2D 1P-3N Boolean Functions}, which form an equivalent set, since the pattern vector of $f_{ i }$ is equivalent to that of its negation $g_{ i }$, $0 \leq i \leq 3$. An  important observation at this point is that, although $f_{ i }$ and $g_{ i }$ are logically different, within our quantum context $f_{ i }$ and $g_{ i }$ are indistinguishable because they lead to the same state. In view of the inability of our classification scheme to distinguish between $f_{ i }$ and its negation $g_{ i }$, we may as well accept this fact. It is very easy to address this issue by performing a second query to the oracle for a single specific input value $\mathbf{ x }$ because the outcome will conclusively differentiate $f_{ i }$ from its negation $g_{ i }$.

For future reference, we gather the Boolean functions $f_{ i }$ into one set, which we call $F_{ 2 }$. Given any function in $F_{ 2 }$, it is easy to construct the corresponding oracle using quantum gates. Accordingly, it is possible to distinguish among the four Boolean functions $f_{ i }$, or, equivalently, among the four $g_{ i }$. Hence, given the promise that the unknown function $f$ is one of the above four Boolean functions, and having the corresponding oracle, the aim of the classification game is to construct a quantum circuit that allows Alice to win with absolute certainty, i.e., with probability $1$ The initial segment of such a circuit is shown in Figure \ref{fig: Initial Segment of The Quantum Circuit for the Classification of f}. $U_{ f }$ is the oracle of the hidden function, chosen by Bob.

\begin{tcolorbox}
	[
		enhanced,
		breakable,
		center title,
		fonttitle = \bfseries,
		grow to left by = 0.500 cm,
		grow to right by = 0.500 cm,
		colback = MagentaLighter!03,
		enhanced jigsaw,			
		sharp corners,
		toprule = 0.001 pt,
		bottomrule = 0.001 pt,
		leftrule = 0.001 pt,
		rightrule = 0.001 pt,
	]
	\begin{minipage} [ b ] { 0.450 \textwidth }
		\begin{figure}[H]
			\centering
			\begin{tikzpicture} [ scale = 1.000 ] 
				\begin{yquant}
					[ operator/separation = 3.000 mm, register/separation = 3.000 mm ]
					qubit { $IR_{ 0 } \colon \ket{ 0 }$ } IR_0;
					qubit { $IR_{ 1 } \colon \ket{ 0 }$ } IR_1;
					qubit { $OR \colon \ket{ - }$ } OR;
					nobit AUX_1;
					[
					name = Input,
					WordBlueDarker,
					line width = 0.250 mm,
					]
					barrier ( - ) ;
					[ draw = WordBlueDarker, fill = WordBlueDarker, radius = 0.400 cm ] box {\color{white} \Large \sf{H}} IR_0;
					[ draw = WordBlueDarker, fill = WordBlueDarker, radius = 0.400 cm ] box {\color{white} \Large \sf{H}} IR_1;
					[
					name = Expansion,
					WordBlueDarker,
					line width = 0.250 mm,
					]
					barrier ( - ) ;
					[ draw = RedPurple, fill = RedPurple, x radius = 0.900 cm, y radius = 0.450 cm ] box { \color{white} \Large \sf{U}$_{ f }$} ( IR_0 - OR );
					[
					name = Output,
					WordBlueDarker,
					line width = 0.250 mm,
					]
					barrier ( - ) ;
					output { $\ket{ \psi_{ 2 } }$ } ( IR_0 - IR_1 );
					\node [ below = 1.750 cm ] at (Input) { $\ket{ \psi_{ 0 } }$ };
					\node [ below = 1.750 cm ] at (Expansion) { $\ket{ \psi_{ 1 } }$ };
				\end{yquant}
				\node [ anchor = center, below = 1.250 cm of Input ] (PhantomNode) { };
			\end{tikzpicture}
			\caption{This figure visualizes the initial segment of a quantum circuit that can be used for the classification of the functions in $F_{ 2 }$.}
			\label{fig: Initial Segment of The Quantum Circuit for the Classification of f}
		\end{figure}
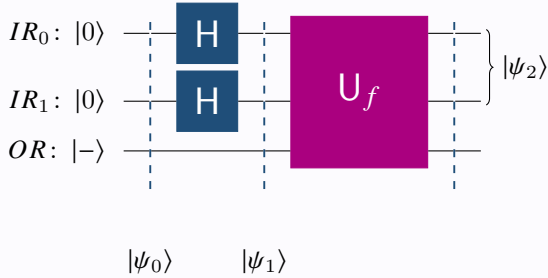
	\end{minipage}
	\hspace{ 0.750 cm }
	\begin{minipage} [ b ] { 0.450 \textwidth }
		\begin{table}[H]
			\caption{The four Boolean functions $f_{ 0 }, f_{ 1 }, f_{ 2 }$ and $f_{ 3 }$ drive the quantum circuit of Figure \ref{fig: Initial Segment of The Quantum Circuit for the Classification of f} to the four different states shown below. In contrast, $f_{ i }$ and $g_{ i }, \ 0 \leq i \leq 3$, are indistinguishable because they lead to the same state.}
			\label{tbl: The State Psi_2 for Each Function of F_2}
			\centering
			\SetTblrInner { rowsep = 1.400 mm }
			\begin{tblr}
				{
					colspec =
					{
						Q [ c, m, 1.500 cm ]
						| [ 1.500 pt, purple7 ]
						| [ 1.500 pt, purple7 ]
						Q [ c, m, 4.750 cm ]
					},
					rowspec =
					{
						|
						[ 3.500 pt, purple4 ]
						|
						[ 0.750 pt, purple4 ]
						|
						[ 0.250 pt, white ]
						Q
						Q
						|
						[ 0.250 pt, white ]
						Q
						|
						[ 0.150 pt, purple4 ]
						Q
						|
						[ 0.150 pt, purple4 ]
						Q
						|
						[ 0.150 pt, purple4 ]
						Q
						|
						[ 0.750 pt, purple4 ]
						|
						[ 0.750 pt, purple4 ]
						|
						[ 0.750 pt, purple4 ]
						Q
						|
						[ 0.150 pt, purple4 ]
						Q
						|
						[ 0.150 pt, purple4 ]
						Q
						|
						[ 0.150 pt, purple4 ]
						Q
						|
						[ 3.500 pt, purple4 ]
					}
				}
				\SetCell [ c = 2 ] { c } The state $\ket{ \psi_{ 2 } }$
				\\
				\SetCell { bg = purple4, fg = white, font = \bfseries } Function
				&
				\SetCell { bg = purple4, fg = white } $\ket{ \pmb{ \psi_{ 2 } } }$
				\\
				$f_{ 0 }$
				&
				$- \frac { 1 } { 2 } \ket{ 00 } + \frac { 1 } { 2 } \ket{ 01 } + \frac { 1 } { 2 } \ket{ 10 } + \frac { 1 } { 2 } \ket{ 11 }$
				\\
				$f_{ 1 }$
				&
				$\phantom{-} \frac { 1 } { 2 } \ket{ 00 } - \frac { 1 } { 2 } \ket{ 01 } + \frac { 1 } { 2 } \ket{ 10 } + \frac { 1 } { 2 } \ket{ 11 }$
				\\
				$f_{ 2 }$
				&
				$\phantom{-} \frac { 1 } { 2 } \ket{ 00 } + \frac { 1 } { 2 } \ket{ 01 } - \frac { 1 } { 2 } \ket{ 10 } + \frac { 1 } { 2 } \ket{ 11 }$
				\\
				$f_{ 3 }$
				&
				$\phantom{-} \frac { 1 } { 2 } \ket{ 00 } + \frac { 1 } { 2 } \ket{ 01 } + \frac { 1 } { 2 } \ket{ 10 } - \frac { 1 } { 2 } \ket{ 11 }$
				\\
				$g_{ 0 }$
				&
				$\phantom{-} \frac { 1 } { 2 } \ket{ 00 } - \frac { 1 } { 2 } \ket{ 01 } - \frac { 1 } { 2 } \ket{ 10 } - \frac { 1 } { 2 } \ket{ 11 }$
				\\
				$g_{ 1 }$
				&
				$- \frac { 1 } { 2 } \ket{ 00 } + \frac { 1 } { 2 } \ket{ 01 } - \frac { 1 } { 2 } \ket{ 10 } - \frac { 1 } { 2 } \ket{ 11 }$
				\\
				$g_{ 2 }$
				&
				$- \frac { 1 } { 2 } \ket{ 00 } - \frac { 1 } { 2 } \ket{ 01 } + \frac { 1 } { 2 } \ket{ 10 } - \frac { 1 } { 2 } \ket{ 11 }$
				\\
				$g_{ 3 }$
				&
				$- \frac { 1 } { 2 } \ket{ 00 } - \frac { 1 } { 2 } \ket{ 01 } - \frac { 1 } { 2 } \ket{ 10 } + \frac { 1 } { 2 } \ket{ 11 }$
			\end{tblr}
		\end{table}
	\end{minipage}
\end{tcolorbox}

Regarding the schematic of Figure \ref{fig: Initial Segment of The Quantum Circuit for the Classification of f}, we note the following.

\begin{itemize}
	\item	
	$IR_{ 0 }$ is the least significant qubit and $IR_{ 1 }$ is the most significant qubit of the quantum input register $IR$ that contains $2$ qubits.
	\item	
	$OR$ is the single-qubit output register that is initialized to state $\ket{ - }$.
	\item	
	$H$ is the Hadamard transform.
	\item	
	$U_{ f }$ is the unitary transform that is based on the oracle for the unknown function $f$ and satisfies relation relation \eqref{eq: Unitary Transform U_f}.
\end{itemize}

After the application of the unitary transform $U_{ f }$, the state of the quantum input register $IR$ will be $\ket{ \psi_{ 2 } }$. As is the norm in such cases, we ignore from now on the output register $OR$ since its state remains $\ket{ - }$. It is quite straightforward to verify the precise dependency of $\ket{ \psi_{ 2 } }$ on each of the functions in $F_{ 2 }$, which is shown in Table \ref{tbl: The State Psi_2 for Each Function of F_2}. The important observation here is that each of the four $f_{ i }$ leads to a different $\ket{ \psi_{ 2 } }$, which means that we can distinguish and classify them. However, as expected, state $\ket{ \psi_{ 2 } }$ is the same for each pair of functions $f_{ i }$ and $g_{ i }$, which means that they are indistinguishable.

Alice now employs a unitary transform that can differentiate among the four Boolean functions $f_{ 0 }, f_{ 1 }, f_{ 2 }$ and $f_{ 3 }$ is $Q_{ 2 }$. The matrix representation of $Q_{ 2 }$ is given by the equation \eqref{eq: Unitary Transform Q_2}. It is easy to verify that the action of $Q_{ 2 }$ on the four possible states $\ket{ \psi_{ 2 } }$ leads to the states shown in Table \ref{tbl: Action of Q_2 on the Functions of F_2}, which are precisely the basis kets of the computational basis $B_{ 4 }$. It is quite straightforward to build $Q_{ 2 }$ using standard quantum gates readily available in contemporary quantum computers. Below we show such a construction that requires only Hadamard, $Z$ and controlled-$Z$ gates:

\begin{align}
	\label{eq: Unitary Transform Q_2}
	Q_{ 2 }
	=
	( H \otimes H )
	\
	CZ
	\
	( Z \otimes Z )
	\
	( H \otimes H )
\end{align}

\begin{tcolorbox}
	[
		enhanced,
		breakable,
		center title,
		fonttitle = \bfseries,
		grow to left by = 0.500 cm,
		grow to right by = 0.500 cm,
		colback = white,			
		enhanced jigsaw,			
		sharp corners,
		toprule = 0.001 pt,
		bottomrule = 0.001 pt,
		leftrule = 0.001 pt,
		rightrule = 0.001 pt,
	]
	\begin{minipage} [ t ] { 0.450 \textwidth }
		\begin{align}
			\label{eq: Unitary Transform Q_2 - Explicit Matrix Form}
			\NiceMatrixOptions{ cell-space-limits = 1.500 pt }
			Q_{ 2 }
			=
			\begin{bNiceMatrix}[ margin ] 
				- \frac { 1 } { 2 } & \phantom{-} \frac { 1 } { 2 } & \phantom{-} \frac { 1 } { 2 } & \phantom{-} \frac { 1 } { 2 } \\
				\phantom{-} \frac { 1 } { 2 } & - \frac { 1 } { 2 } & \phantom{-} \frac { 1 } { 2 } & \phantom{-} \frac { 1 } { 2 } \\
				\phantom{-} \frac { 1 } { 2 } & \phantom{-} \frac { 1 } { 2 } & - \frac { 1 } { 2 } & \phantom{-} \frac { 1 } { 2 } \\
				\phantom{-} \frac { 1 } { 2 } & \phantom{-} \frac { 1 } { 2 } & \phantom{-} \frac { 1 } { 2 } & - \frac { 1 } { 2 } \\
			\end{bNiceMatrix}
		\end{align}
	\end{minipage}
	\hspace{ 0.750 cm }
	\begin{minipage} [ t ] { 0.450 \textwidth }
		\begin{figure}[H]
			\centering
			\includegraphics [ scale = 0.800, trim = {0.000cm 0.750cm 0.000cm 0.500cm}, clip ] {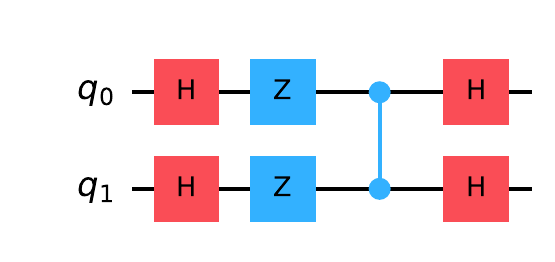}
			\caption{This figure shows the quantum circuit that implements the unitary transform $Q_{ 2 }$ for the classification of the Boolean functions in $F_{ 2 }$.}
			\label{fig: Classifier Kernel Q2}
		\end{figure}
	\end{minipage}
\end{tcolorbox}

\begin{tcolorbox}
	[
		enhanced,
		breakable,
		center title,
		fonttitle = \bfseries,
		grow to left by = 0.000 cm,
		grow to right by = 0.000 cm,
		colback = WordBlueVeryLight!03,
		enhanced jigsaw,			
		sharp corners,
		toprule = 0.001 pt,
		bottomrule = 0.001 pt,
		leftrule = 0.001 pt,
		rightrule = 0.001 pt,
	]
	\begin{table}[H]
		\caption{This table contains the outcome of the action of $Q_{ 2 }$ on the four possible states $\ket{ \psi_{ 2 } }$ outlined in Table \ref{tbl: The State Psi_2 for Each Function of F_2}.}
		\label{tbl: Action of Q_2 on the Functions of F_2}
		\centering
		\SetTblrInner { rowsep = 1.200 mm }
		\begin{tblr}
			{
				colspec =
				{
					Q [ c, m, 3.250 cm ]
					| [ 0.750 pt, azure7 ]
					| [ 0.750 pt, azure7 ]
					Q [ c, m, 2.000 cm ]
					| [ 0.500 pt, azure7 ]
					Q [ c, m, 2.000 cm ]
					| [ 0.500 pt, azure7 ]
					Q [ c, m, 2.000 cm ]
					| [ 0.500 pt, azure7 ]
					Q [ c, m, 2.000 cm ]
				},
				rowspec =
				{
					|
					[ 3.500 pt, azure7 ]
					|
					[ 0.750 pt, azure7 ]
					|
					[ 0.250 pt, white ]
					Q
					|
					Q
					|
					[ 0.500 pt, azure7 ]
					Q
					|
					[ 3.500 pt, azure7 ]
				}
			}
			&
			\SetCell { bg = azure5, fg = white, font = \bfseries } $\mathbf{ f_{ 0 } }$
			&
			\SetCell { bg = azure5, fg = white, font = \bfseries } $\mathbf{ f_{ 1 } }$
			&
			\SetCell { bg = azure5, fg = white, font = \bfseries } $\mathbf{ f_{ 2 } }$
			&
			\SetCell { bg = azure5, fg = white, font = \bfseries } $\mathbf{ f_{ 3 } }$
			\\
			\SetCell { bg = azure5, fg = white, font = \bfseries } $\mathbf{ Q }_{ 2 }$ action on $\ket{ \pmb{ \psi_{ 2 } } }$
			&
			$
			Q_{ 2 }
			\
			\NiceMatrixOptions{cell-space-limits = 1.500 pt}
			\begin{bNiceMatrix}[ margin ]	
				- \frac { 1 } { 2 } \\ \frac { 1 } { 2 } \\ \frac { 1 } { 2 } \\ \frac { 1 } { 2 }
			\end{bNiceMatrix}
			$
			&
			$
			Q_{ 2 }
			\
			\NiceMatrixOptions{cell-space-limits = 1.500 pt}
			\begin{bNiceMatrix}[ margin ]	
				\frac { 1 } { 2 } \\ - \frac { 1 } { 2 } \\ \frac { 1 } { 2 } \\ \frac { 1 } { 2 }
			\end{bNiceMatrix}
			$
			&
			$
			Q_{ 2 }
			\
			\NiceMatrixOptions{cell-space-limits = 1.500 pt}
			\begin{bNiceMatrix}[ margin ]	
				\frac { 1 } { 2 } \\ \frac { 1 } { 2 } \\ - \frac { 1 } { 2 } \\ \frac { 1 } { 2 }
			\end{bNiceMatrix}
			$
			&
			$
			Q_{ 2 }
			\
			\NiceMatrixOptions{cell-space-limits = 1.500 pt}
			\begin{bNiceMatrix}[ margin ]	
				\frac { 1 } { 2 } \\ \frac { 1 } { 2 } \\ \frac { 1 } { 2 } \\ - \frac { 1 } { 2 }
			\end{bNiceMatrix}
			$
			\\
			\SetCell { bg = azure5, fg = white, font = \bfseries } {Outcome}
			&
			$
			\begin{bNiceMatrix}[ margin ]	
				1 \\ 0 \\ 0 \\ 0
			\end{bNiceMatrix}
			=
			\ket{ 00 }
			$
			&
			$
			\begin{bNiceMatrix}[ margin ]	
				0 \\ 1 \\ 0 \\ 0
			\end{bNiceMatrix}
			=
			\ket{ 01 }
			$
			&
			$
			\begin{bNiceMatrix}[ margin ]	
				0 \\ 0 \\ 1 \\ 0
			\end{bNiceMatrix}
			=
			\ket{ 10 }
			$
			&
			$
			\begin{bNiceMatrix}[ margin ]	
				0 \\ 0 \\ 0 \\ 1
			\end{bNiceMatrix}
			=
			\ket{ 11 }
			$
			\\
		\end{tblr}
	\end{table}
\end{tcolorbox}

Therefore, the quantum algorithm that classifies each Boolean function contained in $F_{ 2 }$ can be visualized by the quantum circuit depicted in Figure \ref{fig: The BFPQC Quantum Circuit for $F_{ 2 }$}. Alice surely wins because the action of the classifier $Q_{ 2 }$ results in the final state of the system being one of the four basis kets of the computational basis $B_{ 4 } = \{ \ket{ 00 }, \ket{ 01 }, \ket{ 10 }, \ket{ 11 } \}$. Specifically, if the oracle encodes $f_{ i }$ the final state will be $\ket{ \mathbf{ i } }$, where $\mathbf{ i }$ is the binary representation of the index $i$, $0 \leq i \leq 3$. Therefore, upon the final measurement Alice will surmise the correct hidden function with probability $1$.

\begin{tcolorbox}
	[
		enhanced,
		breakable,
		center title,
		fonttitle = \bfseries,
		grow to left by = 0.000 cm,
		grow to right by = 0.000 cm,
		colback = MagentaLighter!03,
		enhanced jigsaw,			
		sharp corners,
		toprule = 0.001 pt,
		bottomrule = 0.001 pt,
		leftrule = 0.001 pt,
		rightrule = 0.001 pt,
	]
	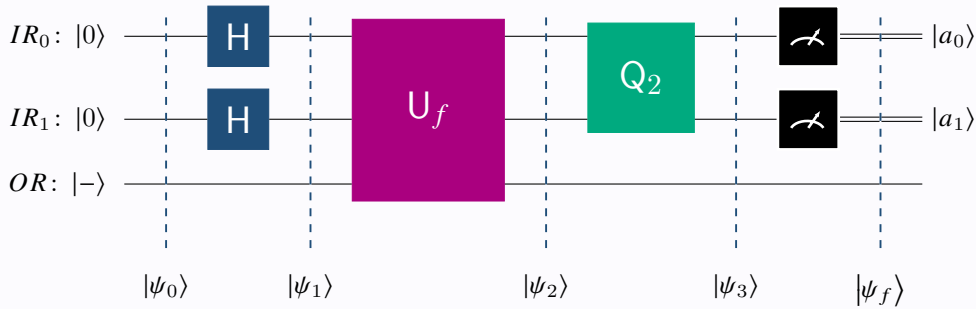
\begin{figure}[H]
		\centering
		\begin{tikzpicture} [ scale = 1.000 ] 
			\begin{yquant}[ operator/separation = 3.000 mm, register/separation = 3.000 mm ]
				qubit { $IR_{ 0 } \colon \ket{ 0 }$ } IR_0;
				qubit { $IR_{ 1 } \colon \ket{ 0 }$ } IR_1;
				qubit { $OR \colon \ket{ - }$ } OR;
				nobit AUX_1;
				[
				name = Input,
				WordBlueDarker,
				line width = 0.250 mm,
				]
				barrier ( - ) ;
				[ draw = WordBlueDarker, fill = WordBlueDarker, radius = 0.400 cm ] box {\color{white} \Large \sf{H}} IR_0;
				[ draw = WordBlueDarker, fill = WordBlueDarker, radius = 0.400 cm ] box {\color{white} \Large \sf{H}} IR_1;
				[
				name = Expansion,
				WordBlueDarker,
				line width = 0.250 mm,
				]
				barrier ( - ) ;
				[ draw = RedPurple, fill = RedPurple, x radius = 1.000 cm, y radius = 0.450 cm ] box { \color{white} \Large \sf{U}$_{ f }$} ( IR_0 - OR );
				[
				name = Oracle,
				WordBlueDarker,
				line width = 0.250 mm,
				]
				barrier ( - ) ;
				[ draw = GreenLighter2, fill = GreenLighter2, x radius = 0.700 cm, y radius = 0.350 cm ] box { \color{white} \Large \sf{Q}$_{ 2 }$}  ( IR_0 - IR_1 );
				[
				name = Classifier,
				WordBlueDarker,
				line width = 0.250 mm,
				]
				barrier ( - ) ;
				[ line width = .350 mm, draw = white, fill = black, radius = 0.400 cm ] measure IR_0;
				[ line width = .350 mm, draw = white, fill = black, radius = 0.400 cm ] measure IR_1;
				[
				name = Measurement,
				WordBlueDarker,
				line width = 0.250 mm,
				]
				barrier ( - ) ;
				output { $\ket{ a_{ 0 } }$ } IR_0;
				output { $\ket{ a_{ 1 } }$ } IR_1;
				\node [ below = 1.750 cm ] at (Input) { $\ket{ \psi_{ 0 } }$ };
				\node [ below = 1.750 cm ] at (Expansion) { $\ket{ \psi_{ 1 } }$ };
				\node [ below = 1.750 cm ] at (Oracle) { $\ket{ \psi_{ 2 } }$ };
				\node [ below = 1.750 cm ] at (Classifier) { $\ket{ \psi_{ 3 } }$ };
				\node [ below = 1.750 cm ] at (Measurement) { $\ket{ \psi_{ f } }$ };
			\end{yquant}
		\end{tikzpicture}
		\caption{This figure visualizes the abstract quantum circuit that implements the BFPQC algorithm for the classification of the functions contained in $F_{ 2 }$.}
		\label{fig: The BFPQC Quantum Circuit for $F_{ 2 }$}
	\end{figure}
\end{tcolorbox}

An actual implementation of the abstract quantum circuit of Figure \ref{fig: The BFPQC Quantum Circuit for $F_{ 2 }$} in Qiskit \cite{Qiskit2025} using the oracle for the function $f_{ 2 }$ is depicted in Figure \ref{fig: Phase4___0100___}. Let us clarify that in all Figures of this paper the qubit numbering follows the ``little-endian'' convention, where is the rightmost qubit is the least significant qubit (LSQ), and the leftmost qubit is the most significant qubit (MSQ).

\begin{tcolorbox}
	[
		enhanced,
		breakable,
		center title,
		fonttitle = \bfseries,
		grow to left by = 0.500 cm,
		grow to right by = 0.500 cm,
		colframe = cyan4,
		colback = white,			
		enhanced jigsaw,			
		sharp corners,
		toprule = 0.000 pt,
		bottomrule = 0.000 pt,
		leftrule = 0.001 pt,
		rightrule = 0.001 pt,
	]
	\begin{figure}[H]
		\centering
		\includegraphics [ scale = 0.520, trim = {0.850cm 0.000cm 0.000cm 0.750cm}, clip ] { "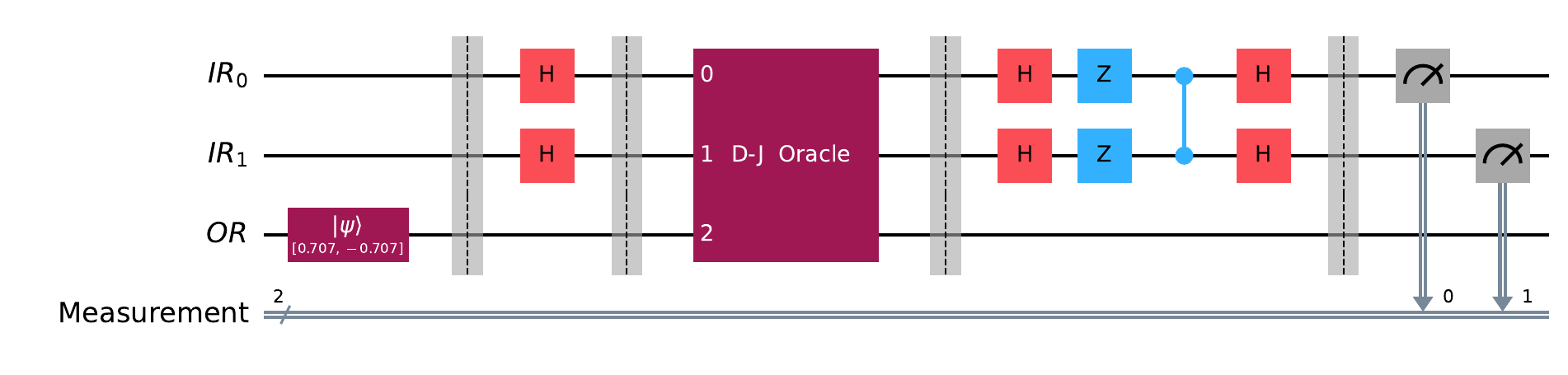" }
		\caption{This figure shows the quantum circuit that implements the BFPQC algorithm for the classification of the Boolean functions in $F_{ 2 }$ using the oracle for $f_{ 2 }$.}
		\label{fig: Phase4___0100___}
	\end{figure}
\end{tcolorbox}

\begin{tcolorbox}
	[
		enhanced,
		breakable,
		center title,
		fonttitle = \bfseries,
		grow to left by = 0.500 cm,
		grow to right by = 0.500 cm,
		colframe = cyan4,
		colback = white,			
		enhanced jigsaw,			
		sharp corners,
		toprule = 0.000 pt,
		bottomrule = 0.000 pt,
		leftrule = 0.001 pt,
		rightrule = 0.001 pt,
	]
	\begin{minipage} [ b ] { 0.475 \textwidth }
		\begin{figure}[H]
			\centering
			\includegraphics [ scale = 0.600, trim = {0.000cm 0.000cm 0.000cm 0.000cm}, clip ] { "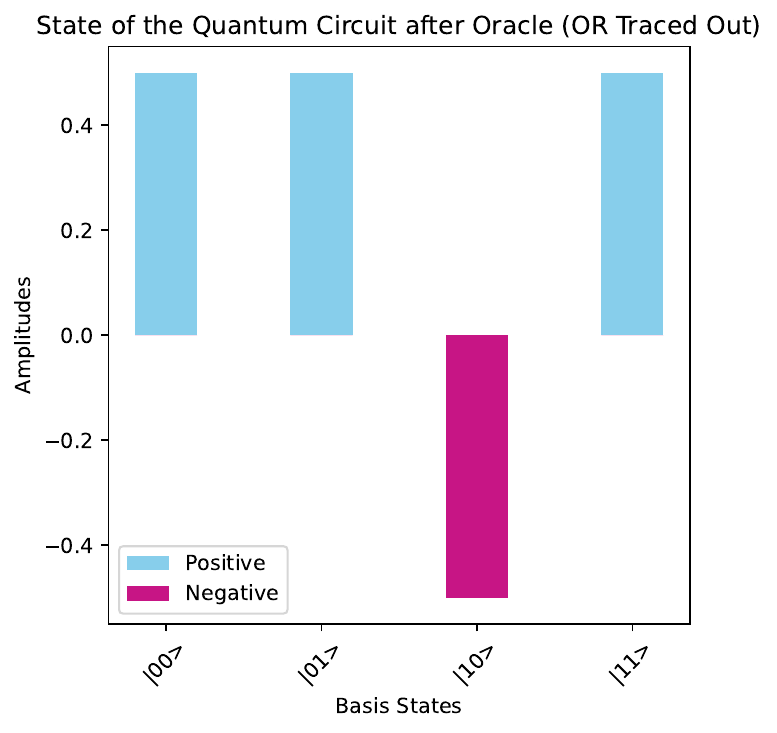" }
			\caption{This is the state of the quantum circuit of Figure \ref{fig: Phase4___0100___} after the oracle but before the action of $Q_{ 2 }$.}
			\label{fig: StateAfterOracle___0100___}
		\end{figure}
	\end{minipage}
	\hspace{ 0.250 cm }
	\begin{minipage} [ b ] { 0.475 \textwidth }
		\begin{figure}[H]
			\centering
			\includegraphics [ scale = 0.600, trim = {0.000cm 0.000cm 0.000cm 0.000cm}, clip ] { "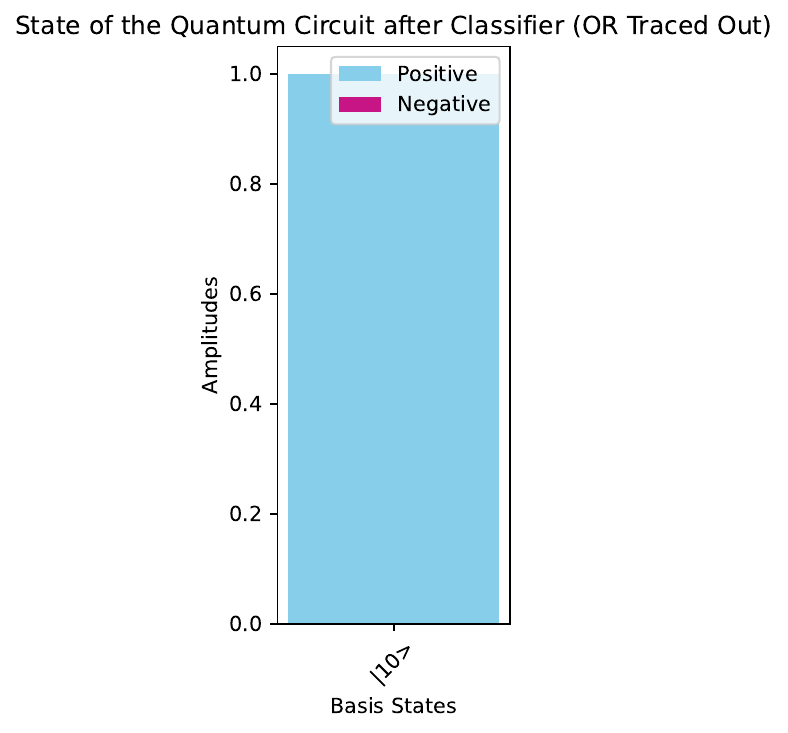" }
			\caption{This is the state of the quantum circuit of Figure \ref{fig: Phase4___0100___} after the action of $Q_{ 2 }$. The subsequent measurement will result in state $\ket{ 10 }$.}
			\label{fig: ___StateAfterQ2___0100___}
		\end{figure}
	\end{minipage}
\end{tcolorbox}
	
\end{example}

\begin{tcolorbox}
	[
		enhanced,
		breakable,
		center title,
		fonttitle = \bfseries,
		colbacktitle = azure4,
		coltitle = white,
		title = The intuition behind the example,
		grow to left by = 0.000 cm,
		grow to right by = 0.000 cm,
		colframe = azure1,
		colback = azure9!75,
		enhanced jigsaw,			
		sharp corners,
		boxrule = 0.500 pt,
	]
	We observe that the four different Boolean functions $f_{ i }$ give rise to four different orthonormal states $\ket{ \psi_{ 2 } }$. Thus, the task of differentiating among the four Boolean functions $f_{ i }$ can be reduced to the task of using a unitary transform that maps the four orthonormal states $\ket{ \psi_{ 2 } }$ to the computational basis $B_{ 4 } = \{ \ket{ 00 }, \ket{ 01 }, \ket{ 10 }, \ket{ 11 } \}$.
\end{tcolorbox}

\section{The general form of the BFPQC algorithm} \label{sec: The General Form of the BFPQC Algorithm}

In this Section we present the general form of the BFPQC algorithm. For this purpose we extend the definitions given in the previous Section.

\begin{definition} {Pattern Basis} { Pattern Basis}
	A \emph{pattern basis} of rank $2 n$, $n \geq 1$, is a collection of $2^{ 2 n }$ pairwise orthogonal pattern vectors of length $2^{ 2 n }$ is denoted by $P_{ 2 n }$.
\end{definition}

The initial pattern basis $P_{ 2 }$ is the set consisting of the following four pairwise orthogonal pattern vectors:

\begin{align}
	\label{eq: Pattern Basis P_2}
	P_{ 2 }
	\coloneq
	\{ 1000, 0100, 0010, 0001 \}
	\ .
\end{align}

Starting from $P_{ 2 }$ we may define an infinite hierarchy of pattern bases. The details are explained below.

\begin{definition} {Pattern Hierarchy} { Pattern Hierarchy}
	We recursively define a \emph{hierarchy} of pattern bases $P_{ 2 n }$, $n \geq 1$, as follows.
	\begin{enumerate}
		[ left = 0.925 cm, labelsep = 0.500 cm, start = 0 ]
		\renewcommand \labelenumi { $($\textbf{PH}$_{ \theenumi }$$)$ }
		\item	If $n = 1$, the corresponding pattern basis is the set $P_{ 2 }$, as defined by \eqref{eq: Pattern Basis P_2}.
		\item	Let $P_{ 2 n }$ contain the pattern vectors $\mathbf{ p }_{ 0 }, \mathbf{ p }_{ 1 }, \dots, \mathbf{ p }_{ m }$; then, the pattern basis $P_{ 2 ( n + 1 ) }$ consists of the pattern vectors with the following syntax structure:
				\begin{align}
					\label{eq: New Pattern Vectors from Old}
					P_{ 2 ( n + 1 ) }
					\coloneq
					\{
					\
					&\overline { \mathbf{ p } }_{ 0 } \mathbf{ p }_{ 0 } \mathbf{ p }_{ 0 } \mathbf{ p }_{ 0 },
					\overline { \mathbf{ p } }_{ 1 } \mathbf{ p }_{ 1 } \mathbf{ p }_{ 1 } \mathbf{ p }_{ 1 },
					\dots,
					\overline { \mathbf{ p } }_{ m } \mathbf{ p }_{ m } \mathbf{ p }_{ m } \mathbf{ p }_{ m },
					\nonumber
					\\
					&\mathbf{ p }_{ 0 } \overline { \mathbf{ p } }_{ 0 } \mathbf{ p }_{ 0 } \mathbf{ p }_{ 0 },
					\mathbf{ p }_{ 1 } \overline { \mathbf{ p } }_{ 1 } \mathbf{ p }_{ 1 } \mathbf{ p }_{ 1 },
					\dots,
					\mathbf{ p }_{ m } \overline { \mathbf{ p } }_{ m } \mathbf{ p }_{ m } \mathbf{ p }_{ m },
					\nonumber
					\\
					&\mathbf{ p }_{ 0 } \mathbf{ p }_{ 0 } \overline { \mathbf{ p } }_{ 0 } \mathbf{ p }_{ 0 },
					\mathbf{ p }_{ 1 } \mathbf{ p }_{ 1 } \overline { \mathbf{ p } }_{ 1 } \mathbf{ p }_{ 1 },
					\dots,
					\mathbf{ p }_{ m } \mathbf{ p }_{ m } \overline { \mathbf{ p } }_{ m } \mathbf{ p }_{ m },
					\nonumber
					\\
					&\mathbf{ p }_{ 0 } \mathbf{ p }_{ 0 } \mathbf{ p }_{ 0 } \overline { \mathbf{ p } }_{ 0 },
					\mathbf{ p }_{ 1 } \mathbf{ p }_{ 1 } \mathbf{ p }_{ 1 } \overline { \mathbf{ p } }_{ 1 },
					\dots,
					\mathbf{ p }_{ m } \mathbf{ p }_{ m } \mathbf{ p }_{ m } \overline { \mathbf{ p } }_{ m }
					\
					\}
				\end{align}
	\end{enumerate}
\end{definition}


An easy conclusion of the above definition is that every $P_{ 2 n }$, $n \geq 1$, contains $2^{ 2 n }$ pairwise orthogonal pattern vectors. Henceforth, we shall assume that the $2^{ 2 n }$ pattern vectors contained in $P_{ 2 n }$ are enumerated as $\mathbf{ p }_{ 0 }, \mathbf{ p }_{ 1 }, \dots, \mathbf{ p }_{ 2^{ 2 n } -1 }$ according to the order prescribed by formula \eqref{eq: New Pattern Vectors from Old}.

\begin{example} {Pattern Basis $P_{ 4 }$} { Pattern Basis $P_{ 4 }$}
To facilitate the understanding of the previous Definition \ref{def: Pattern Hierarchy}, we list $P_{ 4 }$ to show how it is derived from $P_{ 2 } = \{ 1000, 0100, 0010, 0001 \}$.

\begin{tcolorbox}
	[
		enhanced,
		breakable,
		center title,
		fonttitle = \bfseries,
		grow to left by = 0.500 cm,
		grow to right by = 0.500 cm,
		colback = MediumTurquoise!05,
		enhanced jigsaw,			
		sharp corners,
		toprule = 0.001 pt,
		bottomrule = 0.001 pt,
		leftrule = 0.001 pt,
		rightrule = 0.001 pt,
	]
	\begin{table}[H]
		\caption{This table contains the pattern vectors of $P_{ 4 }$.}
		\label{tbl: Pattern Basis $P_{ 4 }$}
		\centering
		\SetTblrInner { rowsep = 1.200 mm }
		\begin{tblr}
			{
				colspec =
				{
					Q [ c, m, 2.350 cm ]
					| [ 0.750 pt, cyan7 ]
					| [ 0.750 pt, cyan7 ]
					Q [ c, m, 2.750 cm ]
					| [ 0.500 pt, cyan7 ]
					Q [ c, m, 2.750 cm ]
					| [ 0.500 pt, cyan7 ]
					Q [ c, m, 2.750 cm ]
					| [ 0.500 pt, cyan7 ]
					Q [ c, m, 2.750 cm ]
				},
				rowspec =
				{
					| [ 3.500 pt, cyan7 ]
					| [ 0.750 pt, cyan7 ]
					| [ 0.250 pt, white ]
					Q
					|
					Q
					| [ 0.150 pt, cyan7 ]
					Q
					| [ 0.150 pt, cyan7 ]
					Q
					| [ 0.150 pt, cyan7 ]
					Q
					| [ 3.500 pt, cyan7 ]
				}
			}
			{ $P_{ 2 }$ \\ Pattern vectors }
			&
			\SetCell { bg = cyan6, fg = white, font = \bfseries } $\overline { \mathbf{ p } }_{ i } \mathbf{ p }_{ i } \mathbf{ p }_{ i } \mathbf{ p }_{ i }$
			&
			\SetCell { bg = cyan6, fg = white, font = \bfseries } $\mathbf{ p }_{ i } \overline { \mathbf{ p } }_{ i } \mathbf{ p }_{ i } \mathbf{ p }_{ i }$
			&
			\SetCell { bg = cyan6, fg = white, font = \bfseries } $\mathbf{ p }_{ i }\mathbf{ p }_{ i } \overline { \mathbf{ p } }_{ i } \mathbf{ p }_{ i }$
			&
			\SetCell { bg = cyan6, fg = white, font = \bfseries } $\mathbf{ p }_{ i } \mathbf{ p }_{ i } \mathbf{ p }_{ i } \overline { \mathbf{ p } }_{ i }$
			\\
			\SetCell { bg = cyan6, fg = white, font = \bfseries } 1000
			&
			{ \footnotesize 0111 \ 1000 \ 1000 \ 1000 }
			&
			{ \footnotesize 1000 \ 0111 \ 1000 \ 1000 }
			&
			{ \footnotesize 1000 \ 1000 \ 0111 \ 1000 }
			&
			{ \footnotesize 1000 \ 1000 \ 1000 \ 0111 }
			\\
			\SetCell { bg = cyan6, fg = white, font = \bfseries } 0100
			&
			{ \footnotesize 1011 \ 0100 \ 0100 \ 0100 }
			&
			{ \footnotesize 0100 \ 1011 \ 0100 \ 0100 }
			&
			{ \footnotesize 0100 \ 0100 \ 1011 \ 0100 }
			&
			{ \footnotesize 0100 \ 0100 \ 0100 \ 1011 }
			\\
			\SetCell { bg = cyan6, fg = white, font = \bfseries } 0010
			&
			{ \footnotesize 1101 \ 0010 \ 0010 \ 0010 }
			&
			{ \footnotesize 0010 \ 1101 \ 0010 \ 0010 }
			&
			{ \footnotesize 0010 \ 0010 \ 1101 \ 0010 }
			&
			{ \footnotesize 0010 \ 0010 \ 0010 \ 1101 }
			\\
			\SetCell { bg = cyan6, fg = white, font = \bfseries } 0001
			&
			{ \footnotesize 1110 \ 0001 \ 0001 \ 0001 }
			&
			{ \footnotesize 0001 \ 1110 \ 0001 \ 0001 }
			&
			{ \footnotesize 0001 \ 0001 \ 1110 \ 0001 }
			&
			{ \footnotesize 0001 \ 0001 \ 0001 \ 1110 }
			\\
		\end{tblr}
	\end{table}
\end{tcolorbox}
	
\end{example}

\begin{definition} {Functions from Patterns} { Functions From Patterns}
	To each pattern basis $P_{ 2 n }$ of rank $2 n$, we associate the class of Boolean functions $f \colon \mathbb{ B }^{ 2 n } \rightarrow \mathbb{ B }$ with the property that their pattern vector is an element of $P_{ 2 n }$.
	We say that this is the class of Boolean functions following the patterns in $P_{ 2 n }$, and we denote it by $F_{ 2 n }$.
\end{definition}

Hence, a hierarchy $P_{ 2 n }$ of pattern bases induces a corresponding hierarchy $F_{ 2 n }$ of classes of Boolean functions. In what follows we shall also assume that the $2^{ 2 n }$ Boolean functions contained in $F_{ 2 n }$ are enumerated as $f_{ 0 }, f_{ 1 }, \dots, f_{ 2^{ 2 n } -1 }$ following the same enumeration with the pattern vectors of $P_{ 2 n }$.

By construction, the pattern vectors, and, consequently, the pattern bases, satisfy the following important relations.

\begin{enumerate}
	[ left = 0.600 cm, labelsep = 0.500 cm, start = 1 ]
	\renewcommand \labelenumi { $($\textbf{R}$_{ \theenumi }$$)$ }
	\item	As we have mentioned in Example \ref{xmp: A Toy Scale Example - 1}, the imbalance ratio $\rho$ of $P_{ 2 } = \{ 1000$, $0100$, $0010$, $0001 \}$ is $\rho = \frac { 1 } { 4 }$.
	\item	The recursive Definition \ref{def: Pattern Hierarchy} of the pattern hierarchy implies that the imbalance ratio satisfies the recurrence relation given below
			\begin{align}
				\label{eq: Imbalance Ratio Recurrence Relation}
				\rho_{ 2 n }
				=
				\frac { 1 } { 4 }
				+
				\frac { 1 } { 2 }
				\
				\rho_{ 2 n  - 2 }
				\qquad
				( n \geq 2 )
				\ ,
			\end{align}
	where $\rho_{ 2 n  - 2 }$ and $\rho_{ 2 n }$ are the imbalance ratios of $P_{ 2 n  - 2 }$ and $P_{ 2 n }$, respectively.
	\item	After some manipulation, the above recurrence relation can be transformed into the next closed form
			\begin{align}
				\label{eq: Closed Form Recurrence Relation}
				\rho_{ 2 n }
				=
				\frac { 1 } { 2 }
				-
				\frac { 1 } { 2^{ n + 1 } }
				\qquad
				( n \geq 1 )
				\ .
			\end{align}
	\item	The above closed form formula enables us to surmise that
			\begin{align}
				\label{eq: Imbalance ratio Main Property}
				\rho_{ 2 n }
				<
				\frac { 1 } { 2 }
				\qquad
				( n \geq 1 )
				\ ,
			\end{align}
	which proves that every pattern basis and every class of Boolean functions in their respective hierarchies have imbalance ratio $< \frac { 1 } { 2 }$, or, in simpler terms, all the Boolean functions we classify are indeed imbalanced as we have previously asserted.
\end{enumerate}

Our purpose is to realize the Boolean Function Pattern Quantum Classifier algorithm through a family of quantum circuits denoted by QCPC$_{ 2 n }$, $n \geq 1$, such that QCPC$_{ 2 n }$ classifies the class of Boolean functions $F_{ 2 n }$, which consist of functions that follow the motif prescribed by the elements of the pattern basis $B_{ 2 n }$. In these quantum circuits, the critical component for the classification is the $Q_{ 2 n }$ unitary classifier, defined below.

\begin{definition} {A Hierarchy of Unitary Classifiers} { A Hierarchy of Unitary Classifiers}
	We recursively define a \emph{hierarchy} of unitary classifiers, denoted by $Q_{ 2 n }$, $n \geq 1$, as follows.
	\begin{enumerate}
		[ left = 0.925 cm, labelsep = 0.500 cm, start = 0 ]
		\renewcommand \labelenumi { $($\textbf{QH}$_{ \theenumi }$$)$ }
		\item	If $n = 1$, the corresponding classifier is $Q_{ 2 }$, as expressed by \eqref{eq: Unitary Transform Q_2} with the matrix representation given by \eqref{fig: Classifier Kernel Q2}.
		\item	Given $Q_{ 2 n }$, the classifier $Q_{ 2 ( n + 1 ) }$ is defined as
				\begin{align}
					\label{eq: New Classifiers from Old}
					Q_{ 2 ( n + 1 ) }
					\coloneq
					Q_{ 2 }
					\otimes
					Q_{ 2 n }
					=
					Q_{ 2 }^{ \otimes ( n + 1 ) }
					\qquad
					( n \geq 1 )
					\ .
				\end{align}
	\end{enumerate}
\end{definition}

\begin{example} {Unitary Classifier $Q_{ 4 }$} { Unitary Classifier $Q_{ 4 }$}
It is instructive to show in detail how the matrix representation of the unitary classifier $Q_{ 4 }$ is derived. By Definition \ref{def: A Hierarchy of Unitary Classifiers}, we know that

\begin{align}
	\label{eq: Unitary Transform Q_4 - Explicit Matrix Form}
	\hspace{ - 2.000 cm }
	Q_{ 4 }
	&\overset { \eqref{eq: New Classifiers from Old} } { = }
	Q_{ 2 }
	\otimes
	Q_{ 2 }
	\overset { \eqref{fig: Classifier Kernel Q2} } { = }
	\NiceMatrixOptions{cell-space-limits = 1.500 pt}
	\begin{bNiceMatrix}[ margin ] 
		- \frac { 1 } { 2 } & \phantom{-} \frac { 1 } { 2 } & \phantom{-} \frac { 1 } { 2 } & \phantom{-} \frac { 1 } { 2 } \\
		\phantom{-} \frac { 1 } { 2 } & - \frac { 1 } { 2 } & \phantom{-} \frac { 1 } { 2 } & \phantom{-} \frac { 1 } { 2 } \\
		\phantom{-} \frac { 1 } { 2 } & \phantom{-} \frac { 1 } { 2 } & - \frac { 1 } { 2 } & \phantom{-} \frac { 1 } { 2 } \\
		\phantom{-} \frac { 1 } { 2 } & \phantom{-} \frac { 1 } { 2 } & \phantom{-} \frac { 1 } { 2 } & - \frac { 1 } { 2 } \\
	\end{bNiceMatrix}
	\otimes
	\begin{bNiceMatrix}[ margin ] 
		- \frac { 1 } { 2 } & \phantom{-} \frac { 1 } { 2 } & \phantom{-} \frac { 1 } { 2 } & \phantom{-} \frac { 1 } { 2 } \\
		\phantom{-} \frac { 1 } { 2 } & - \frac { 1 } { 2 } & \phantom{-} \frac { 1 } { 2 } & \phantom{-} \frac { 1 } { 2 } \\
		\phantom{-} \frac { 1 } { 2 } & \phantom{-} \frac { 1 } { 2 } & - \frac { 1 } { 2 } & \phantom{-} \frac { 1 } { 2 } \\
		\phantom{-} \frac { 1 } { 2 } & \phantom{-} \frac { 1 } { 2 } & \phantom{-} \frac { 1 } { 2 } & - \frac { 1 } { 2 } \\
	\end{bNiceMatrix}
	\nonumber
	\\
	&\hspace { 0.100 cm } =
	\begin{bNiceMatrix}[ margin, cell-space-limits = 2.000 pt ] 
		- \frac { 1 } { 2 } Q_{ 2 } & \phantom{-} \frac { 1 } { 2 } Q_{ 2 } &
		\phantom{-} \frac { 1 } { 2 } Q_{ 2 } & \phantom{-} \frac { 1 } { 2 } Q_{ 2 }
		\\
		\phantom{-} \frac { 1 } { 2 } Q_{ 2 } & - \frac { 1 } { 2 } Q_{ 2 } &
		\phantom{-} \frac { 1 } { 2 } Q_{ 2 } & \phantom{-} \frac { 1 } { 2 } Q_{ 2 }
		\\
		\phantom{-} \frac { 1 } { 2 } Q_{ 2 } & \phantom{-} \frac { 1 } { 2 } Q_{ 2 } &
		- \frac { 1 } { 2 } Q_{ 2 } & \phantom{-} \frac { 1 } { 2 } Q_{ 2 }
		\\
		\phantom{-} \frac { 1 } { 2 } Q_{ 2 } & \phantom{-} \frac { 1 } { 2 } Q_{ 2 } &
		\phantom{-} \frac { 1 } { 2 } Q_{ 2 } & - \frac { 1 } { 2 } Q_{ 2 }
		\CodeAfter
		\tikz \draw
		[
		purple4,
		line width = 0.400 mm,
		shorten <= 2.000 mm,
		shorten >= 2.000 mm,
		]
		(3 -| 1) -- (3 -| last);
		%
		\tikz \draw
		[
		purple4,
		line width = 0.400 mm,
		shorten <= 2.000 mm,
		]
		(1 -| 3) -- (last -| 3);
		%
		\tikz \draw
		[
		teal7 ,
		line width = 0.200 mm,
		shorten <= 3.000 mm,
		shorten >= 3.000 mm,
		]
		(2 -| 1) -- (2 -| 3);
		%
		\tikz \draw
		[
		teal7 ,
		line width = 0.200 mm,
		shorten <= 3.000 mm,
		shorten >= 3.000 mm,
		]
		(2 -| 3) -- (2 -| last);
		%
		\tikz \draw
		[
		teal7 ,
		line width = 0.200 mm,
		shorten <= 3.000 mm,
		shorten >= 3.000 mm,
		]
		(4 -| 1) -- (4 -| 3);
		%
		\tikz \draw
		[
		teal7 ,
		line width = 0.200 mm,
		shorten <= 3.000 mm,
		shorten >= 3.000 mm,
		]
		(4 -| 3) -- (4 -| last);
		%
		\tikz \draw
		[
		teal7,
		line width = 0.200 mm,
		shorten <= 2.000 mm,
		]
		(1 -| 2) -- (3 -| 2);
		%
		\tikz \draw
		[
		teal7,
		line width = 0.200 mm,
		shorten <= 2.000 mm,
		]
		(1 -| 4) -- (3 -| 4);
		%
		\tikz \draw
		[
		teal7,
		line width = 0.200 mm,
		shorten <= 2.000 mm,
		]
		(3 -| 2) -- (last -| 2);
		%
		\tikz \draw
		[
		teal7,
		line width = 0.200 mm,
		shorten <= 2.000 mm,
		]
		(3 -| 4) -- (last -| 4);
	\end{bNiceMatrix}
	\nonumber
	\\
	&\hspace { 0.100 cm } =
	{\footnotesize
		\begin{bNiceMatrix}[ margin, cell-space-limits = 2.000 pt ] 
			\noalign{\vskip 3pt}
			\phantom{-} \frac { 1 } { 4 } & - \frac { 1 } { 4 } & - \frac { 1 } { 4 } & - \frac { 1 } { 4 } &
			- \frac { 1 } { 4 } & \phantom{-} \frac { 1 } { 4 } & \phantom{-} \frac { 1 } { 4 } & \phantom{-} \frac { 1 } { 4 } &
			- \frac { 1 } { 4 } & \phantom{-} \frac { 1 } { 4 } & \phantom{-} \frac { 1 } { 4 } & \phantom{-} \frac { 1 } { 4 } &
			- \frac { 1 } { 4 } & \phantom{-} \frac { 1 } { 4 } & \phantom{-} \frac { 1 } { 4 } & \phantom{-} \frac { 1 } { 4 }
			\\
			- \frac { 1 } { 4 } & \phantom{-} \frac { 1 } { 4 } & - \frac { 1 } { 4 } & - \frac { 1 } { 4 } &
			\phantom{-} \frac { 1 } { 4 } & - \frac { 1 } { 4 } & \phantom{-} \frac { 1 } { 4 } & \phantom{-} \frac { 1 } { 4 } &
			\phantom{-} \frac { 1 } { 4 } & - \frac { 1 } { 4 } & \phantom{-} \frac { 1 } { 4 } & \phantom{-} \frac { 1 } { 4 } &
			\phantom{-} \frac { 1 } { 4 } & - \frac { 1 } { 4 } & \phantom{-} \frac { 1 } { 4 } & \phantom{-} \frac { 1 } { 4 }
			\\
			- \frac { 1 } { 4 } & - \frac { 1 } { 4 } & \phantom{-} \frac { 1 } { 4 } & - \frac { 1 } { 4 } &
			\phantom{-} \frac { 1 } { 4 } & \phantom{-} \frac { 1 } { 4 } & - \frac { 1 } { 4 } & \phantom{-} \frac { 1 } { 4 } &
			\phantom{-} \frac { 1 } { 4 } & \phantom{-} \frac { 1 } { 4 } & - \frac { 1 } { 4 } & \phantom{-} \frac { 1 } { 4 } &
			\phantom{-} \frac { 1 } { 4 } & \phantom{-} \frac { 1 } { 4 } & - \frac { 1 } { 4 } & \phantom{-} \frac { 1 } { 4 }
			\\
			- \frac { 1 } { 4 } & - \frac { 1 } { 4 } & - \frac { 1 } { 4 } & \phantom{-} \frac { 1 } { 4 } &
			\phantom{-} \frac { 1 } { 4 } & \phantom{-} \frac { 1 } { 4 } & \phantom{-} \frac { 1 } { 4 } & - \frac { 1 } { 4 } &
			\phantom{-} \frac { 1 } { 4 } & \phantom{-} \frac { 1 } { 4 } & \phantom{-} \frac { 1 } { 4 } & - \frac { 1 } { 4 } &
			\phantom{-} \frac { 1 } { 4 } & \phantom{-} \frac { 1 } { 4 } & \phantom{-} \frac { 1 } { 4 } & - \frac { 1 } { 4 }
			\\
			- \frac { 1 } { 4 } & \phantom{-} \frac { 1 } { 4 } & \phantom{-} \frac { 1 } { 4 } & \phantom{-} \frac { 1 } { 4 } &
			\phantom{-} \frac { 1 } { 4 } & - \frac { 1 } { 4 } & - \frac { 1 } { 4 } & - \frac { 1 } { 4 } &
			- \frac { 1 } { 4 } & \phantom{-} \frac { 1 } { 4 } & \phantom{-} \frac { 1 } { 4 } & \phantom{-} \frac { 1 } { 4 } &
			- \frac { 1 } { 4 } & \phantom{-} \frac { 1 } { 4 } & \phantom{-} \frac { 1 } { 4 } & \phantom{-} \frac { 1 } { 4 }
			\\
			\phantom{-} \frac { 1 } { 4 } & - \frac { 1 } { 4 } & \phantom{-} \frac { 1 } { 4 } & \phantom{-} \frac { 1 } { 4 } &
			- \frac { 1 } { 4 } & \phantom{-} \frac { 1 } { 4 } & - \frac { 1 } { 4 } & - \frac { 1 } { 4 } &
			\phantom{-} \frac { 1 } { 4 } & - \frac { 1 } { 4 } & \phantom{-} \frac { 1 } { 4 } & \phantom{-} \frac { 1 } { 4 } &
			\phantom{-} \frac { 1 } { 4 } & - \frac { 1 } { 4 } & \phantom{-} \frac { 1 } { 4 } & \phantom{-} \frac { 1 } { 4 }
			\\
			\phantom{-} \frac { 1 } { 4 } & \phantom{-} \frac { 1 } { 4 } & - \frac { 1 } { 4 } & \phantom{-} \frac { 1 } { 4 } &
			- \frac { 1 } { 4 } & - \frac { 1 } { 4 } & \phantom{-} \frac { 1 } { 4 } & - \frac { 1 } { 4 } &
			\phantom{-} \frac { 1 } { 4 } & \phantom{-} \frac { 1 } { 4 } & - \frac { 1 } { 4 } & \phantom{-} \frac { 1 } { 4 } &
			\phantom{-} \frac { 1 } { 4 } & \phantom{-} \frac { 1 } { 4 } & - \frac { 1 } { 4 } & \phantom{-} \frac { 1 } { 4 }
			\\
			\phantom{-} \frac { 1 } { 4 } & \phantom{-} \frac { 1 } { 4 } & \phantom{-} \frac { 1 } { 4 } & - \frac { 1 } { 4 } &
			- \frac { 1 } { 4 } & - \frac { 1 } { 4 } & - \frac { 1 } { 4 } & \phantom{-} \frac { 1 } { 4 } &
			\phantom{-} \frac { 1 } { 4 } & \phantom{-} \frac { 1 } { 4 } & \phantom{-} \frac { 1 } { 4 } & - \frac { 1 } { 4 } &
			\phantom{-} \frac { 1 } { 4 } & \phantom{-} \frac { 1 } { 4 } & \phantom{-} \frac { 1 } { 4 } & - \frac { 1 } { 4 }
			\\
			- \frac { 1 } { 4 } & \phantom{-} \frac { 1 } { 4 } & \phantom{-} \frac { 1 } { 4 } & \phantom{-} \frac { 1 } { 4 } &
			- \frac { 1 } { 4 } & \phantom{-} \frac { 1 } { 4 } & \phantom{-} \frac { 1 } { 4 } & \phantom{-} \frac { 1 } { 4 } &
			\phantom{-} \frac { 1 } { 4 } & - \frac { 1 } { 4 } & - \frac { 1 } { 4 } & - \frac { 1 } { 4 } &
			- \frac { 1 } { 4 } & \phantom{-} \frac { 1 } { 4 } & \phantom{-} \frac { 1 } { 4 } & \phantom{-} \frac { 1 } { 4 }
			\\
			\phantom{-} \frac { 1 } { 4 } & - \frac { 1 } { 4 } & \phantom{-} \frac { 1 } { 4 } & \phantom{-} \frac { 1 } { 4 } &
			\phantom{-} \frac { 1 } { 4 } & - \frac { 1 } { 4 } & \phantom{-} \frac { 1 } { 4 } & \phantom{-} \frac { 1 } { 4 } &
			- \frac { 1 } { 4 } & \phantom{-} \frac { 1 } { 4 } & - \frac { 1 } { 4 } & - \frac { 1 } { 4 } &
			\phantom{-} \frac { 1 } { 4 } & - \frac { 1 } { 4 } & \phantom{-} \frac { 1 } { 4 } & \phantom{-} \frac { 1 } { 4 }
			\\
			\phantom{-} \frac { 1 } { 4 } & \phantom{-} \frac { 1 } { 4 } & - \frac { 1 } { 4 } & \phantom{-} \frac { 1 } { 4 } &
			\phantom{-} \frac { 1 } { 4 } & \phantom{-} \frac { 1 } { 4 } & - \frac { 1 } { 4 } & \phantom{-} \frac { 1 } { 4 } &
			- \frac { 1 } { 4 } & - \frac { 1 } { 4 } & \phantom{-} \frac { 1 } { 4 } & - \frac { 1 } { 4 } &
			\phantom{-} \frac { 1 } { 4 } & \phantom{-} \frac { 1 } { 4 } & - \frac { 1 } { 4 } & \phantom{-} \frac { 1 } { 4 }
			\\
			\phantom{-} \frac { 1 } { 4 } & \phantom{-} \frac { 1 } { 4 } & \phantom{-} \frac { 1 } { 4 } & - \frac { 1 } { 4 } &
			\phantom{-} \frac { 1 } { 4 } & \phantom{-} \frac { 1 } { 4 } & \phantom{-} \frac { 1 } { 4 } & - \frac { 1 } { 4 } &
			- \frac { 1 } { 4 } & - \frac { 1 } { 4 } & - \frac { 1 } { 4 } & \phantom{-} \frac { 1 } { 4 } &
			\phantom{-} \frac { 1 } { 4 } & \phantom{-} \frac { 1 } { 4 } & \phantom{-} \frac { 1 } { 4 } & - \frac { 1 } { 4 }
			\\
			- \frac { 1 } { 4 } & \phantom{-} \frac { 1 } { 4 } & \phantom{-} \frac { 1 } { 4 } & \phantom{-} \frac { 1 } { 4 } &
			- \frac { 1 } { 4 } & \phantom{-} \frac { 1 } { 4 } & \phantom{-} \frac { 1 } { 4 } & \phantom{-} \frac { 1 } { 4 } &
			- \frac { 1 } { 4 } & \phantom{-} \frac { 1 } { 4 } & \phantom{-} \frac { 1 } { 4 } & \phantom{-} \frac { 1 } { 4 } &
			\phantom{-} \frac { 1 } { 4 } & - \frac { 1 } { 4 } & - \frac { 1 } { 4 } & - \frac { 1 } { 4 }
			\\
			\phantom{-} \frac { 1 } { 4 } & - \frac { 1 } { 4 } & \phantom{-} \frac { 1 } { 4 } & \phantom{-} \frac { 1 } { 4 } &
			\phantom{-} \frac { 1 } { 4 } & - \frac { 1 } { 4 } & \phantom{-} \frac { 1 } { 4 } & \phantom{-} \frac { 1 } { 4 } &
			\phantom{-} \frac { 1 } { 4 } & - \frac { 1 } { 4 } & \phantom{-} \frac { 1 } { 4 } & \phantom{-} \frac { 1 } { 4 } &
			- \frac { 1 } { 4 } & \phantom{-} \frac { 1 } { 4 } & - \frac { 1 } { 4 } & - \frac { 1 } { 4 }
			\\
			\phantom{-} \frac { 1 } { 4 } & \phantom{-} \frac { 1 } { 4 } & - \frac { 1 } { 4 } & \phantom{-} \frac { 1 } { 4 } &
			\phantom{-} \frac { 1 } { 4 } & \phantom{-} \frac { 1 } { 4 } & - \frac { 1 } { 4 } & \phantom{-} \frac { 1 } { 4 } &
			\phantom{-} \frac { 1 } { 4 } & \phantom{-} \frac { 1 } { 4 } & - \frac { 1 } { 4 } & \phantom{-} \frac { 1 } { 4 } &
			- \frac { 1 } { 4 } & - \frac { 1 } { 4 } & \phantom{-} \frac { 1 } { 4 } & - \frac { 1 } { 4 }
			\\
			\phantom{-} \frac { 1 } { 4 } & \phantom{-} \frac { 1 } { 4 } & \phantom{-} \frac { 1 } { 4 } & - \frac { 1 } { 4 } &
			\phantom{-} \frac { 1 } { 4 } & \phantom{-} \frac { 1 } { 4 } & \phantom{-} \frac { 1 } { 4 } & - \frac { 1 } { 4 } &
			\phantom{-} \frac { 1 } { 4 } & \phantom{-} \frac { 1 } { 4 } & \phantom{-} \frac { 1 } { 4 } & - \frac { 1 } { 4 } &
			- \frac { 1 } { 4 } & - \frac { 1 } { 4 } & - \frac { 1 } { 4 } & \phantom{-} \frac { 1 } { 4 }
			\CodeAfter
			\tikz \draw
			[
			purple4,
			line width = 0.400 mm,
			shorten <= 2.000 mm,
			shorten >= 2.000 mm,
			]
			(9 -| 1) -- (9 -| last);
			\tikz \draw
			[
			purple4,
			line width = 0.400 mm,
			shorten <= 2.000 mm,
			]
			(1 -| 9) -- (last -| 9);
			\tikz \draw
			[
			teal7 ,
			line width = 0.250 mm,
			shorten <= 3.000 mm,
			shorten >= 3.000 mm,
			]
			(5 -| 1) -- (5 -| 9);
			\tikz \draw
			[
			teal7 ,
			line width = 0.250 mm,
			shorten <= 3.000 mm,
			shorten >= 3.000 mm,
			]
			(5 -| 9) -- (5 -| last);
			\tikz \draw
			[
			teal7 ,
			line width = 0.250 mm,
			shorten <= 3.000 mm,
			shorten >= 3.000 mm,
			]
			(13 -| 1) -- (13 -| 9);
			\tikz \draw
			[
			teal7 ,
			line width = 0.250 mm,
			shorten <= 3.000 mm,
			shorten >= 3.000 mm,
			]
			(13 -| 9) -- (13 -| last);
			\tikz \draw
			[
			teal7 ,
			line width = 0.250 mm,
			shorten <= 3.000 mm,
			shorten >= 3.000 mm,
			]
			(1 -| 5) -- (9 -| 5);
			\tikz \draw
			[
			teal7 ,
			line width = 0.250 mm,
			shorten <= 3.000 mm,
			shorten >= 3.000 mm,
			]
			(9 -| 5) -- (last -| 5);
			\tikz \draw
			[
			teal7 ,
			line width = 0.250 mm,
			shorten <= 3.000 mm,
			shorten >= 3.000 mm,
			]
			(1 -| 13) -- (9 -| 13);
			\tikz \draw
			[
			teal7 ,
			line width = 0.250 mm,
			shorten <= 3.000 mm,
			shorten >= 3.000 mm,
			]
			(9 -| 13) -- (last -| 13);
		\end{bNiceMatrix}
	}
	\ .
\end{align}

\end{example}

We use the unitary classifier $Q_{ 2 n }$, $n \geq 1$, as the main component in the construction of the family of quantum circuits QCPC$_{ 2 n }$, for the classification of the the class of Boolean functions $F_{ 2 n }$.

\begin{definition} {A Family of Quantum Classifiers} { A Family of Quantum Classifiers}
	To each unitary classifier $Q_{ 2 n }, n \geq 1$, we associate the quantum circuits QCPC$_{ 2 n }$, for the classification of the class of Boolean functions $F_{ 2 n }$.

	\begin{itemize}
		\item	
		The first member of this family, the QCPC$_{ 2 }$ quantum circuit, takes the form depicted in Figure \ref{fig: The BFPQC Quantum Circuit for $F_{ 2 }$} and can classify the Boolean functions in $F_{ 2 }$.
		\item	
		The general QCPC$_{ 2 n }$ quantum circuit takes the abstract form visualized in Figure \ref{fig: The BFPQC Quantum Circuit for $F_{ 2 n }$}. It is endowed with the oracle $U_{ f} $ encoding the behavior of the Boolean function $f$, which is promised to belong to $F_{ 2 n }$.
	\end{itemize}

\end{definition}

\begin{tcolorbox}
	[
		enhanced,
		breakable,
		center title,
		fonttitle = \bfseries,
		grow to left by = 0.000 cm,
		grow to right by = 0.000 cm,
		colback = MagentaLighter!03,
		enhanced jigsaw,			
		sharp corners,
		toprule = 0.001 pt,
		bottomrule = 0.001 pt,
		leftrule = 0.001 pt,
		rightrule = 0.001 pt,
	]
	\begin{figure}[H]
		\centering
		\begin{tikzpicture} [ scale = 1.000 ] 
			\begin{yquant}[ operator/separation = 3.000 mm, register/separation = 3.000 mm, every nobit output/.style = { } ]
				qubit { $IR_{ 0 } \colon \ket{ 0 }$ } IR;
				qubit { $IR_{ 1 } \colon \ket{ 0 }$ } IR [ + 1 ];
				qubit { $IR_{ 2 } \colon \ket{ 0 }$ } IR [ + 1 ];
				qubit { $IR_{ 3 } \colon \ket{ 0 }$ } IR [ + 1 ];
				qubit { $\vdots$ \hspace{ 0.475 cm } } IR [ + 1 ]; discard IR [ 4 ];
				qubit { $IR_{ 2 n - 2 } \colon \ket{ 0 }$ } IR [ + 1 ];
				qubit { $IR_{ 2 n - 1 } \colon \ket{ 0 }$ } IR [ + 1 ];
				qubit { $OR \colon \ket{ - }$ } OR;
				nobit AUX_1;
				[
				name = Input,
				WordBlueDarker,
				line width = 0.250 mm,
				]
				barrier ( - ) ;
				[ draw = WordBlueDarker, fill = WordBlueDarker, radius = 0.400 cm ] box {\color{white} \Large \sf{H}} IR [ 0 ];
				[ draw = WordBlueDarker, fill = WordBlueDarker, radius = 0.400 cm ] box {\color{white} \Large \sf{H}} IR [ 1 ];
				[ draw = WordBlueDarker, fill = WordBlueDarker, radius = 0.400 cm ] box {\color{white} \Large \sf{H}} IR [ 2 ];
				[ draw = WordBlueDarker, fill = WordBlueDarker, radius = 0.400 cm ] box {\color{white} \Large \sf{H}} IR [ 3 ];
				[ draw = WordBlueDarker, fill = WordBlueDarker, radius = 0.400 cm ] box {\color{white} \Large \sf{H}} IR [ 5 ];
				[ draw = WordBlueDarker, fill = WordBlueDarker, radius = 0.400 cm ] box {\color{white} \Large \sf{H}} IR [ 6 ];
				[
				name = Expansion,
				WordBlueDarker,
				line width = 0.250 mm,
				]
				barrier ( - ) ;
				[ draw = RedPurple, fill = RedPurple, x radius = 0.900 cm, y radius = 0.450 cm ] box { \color{white} \Large \sf{U}$_{ f }$} ( IR - OR );
				[
				name = Oracle,
				WordBlueDarker,
				line width = 0.250 mm,
				]
				barrier ( - ) ;
				[ draw = GreenLighter2, fill = GreenLighter2, x radius = 0.700 cm, y radius = 0.350 cm ] box { \color{white} \Large \sf{Q}$_{ 2 }$}  ( IR [ 0 ] - IR [ 1 ] );
				[ draw = GreenLighter2, fill = GreenLighter2, x radius = 0.700 cm, y radius = 0.350 cm ] box { \color{white} \Large \sf{Q}$_{ 2 }$}  ( IR [ 2 ] - IR [ 3 ] );
				[ draw = GreenLighter2, fill = GreenLighter2, x radius = 0.700 cm, y radius = 0.350 cm ] box { \color{white} \Large \sf{Q}$_{ 2 }$}  ( IR [ 5 ] - IR [ 6 ] );
				[
				name = Classifier,
				WordBlueDarker,
				line width = 0.250 mm,
				]
				barrier ( - ) ;
				[ line width = .350 mm, draw = white, fill = black, radius = 0.400 cm ] measure IR [ 0 ];
				[ line width = .350 mm, draw = white, fill = black, radius = 0.400 cm ] measure IR [ 1 ];
				[ line width = .350 mm, draw = white, fill = black, radius = 0.400 cm ] measure IR [ 2 ];
				[ line width = .350 mm, draw = white, fill = black, radius = 0.400 cm ] measure IR [ 3 ];
				[ line width = .350 mm, draw = white, fill = black, radius = 0.400 cm ] measure IR [ 5 ];
				[ line width = .350 mm, draw = white, fill = black, radius = 0.400 cm ] measure IR [ 6 ];
				[
				name = Measurement,
				WordBlueDarker,
				line width = 0.250 mm,
				]
				barrier ( - ) ;
				output { $\ket{ a_{ 0 } }$ } IR [ 0 ];
				output { $\ket{ a_{ 1 } }$ } IR [ 1 ];
				output { $\ket{ a_{ 2 } }$ } IR [ 2 ];
				output { $\ket{ a_{ 3 } }$ } IR [ 3 ];
				output { $\vdots$ } IR [ 4 ];
				output { $\ket{ a_{ 2 n - 2 } }$ } IR [ 5 ];
				output { $\ket{ a_{ 2 n - 1 } }$ } IR [ 6 ];
				\node [ below = 4.500 cm ] at (Input) { $\ket{ \psi_{ 0 } }$ };
				\node [ below = 4.500 cm ] at (Expansion) { $\ket{ \psi_{ 1 } }$ };
				\node [ below = 4.500 cm ] at (Oracle) { $\ket{ \psi_{ 2 } }$ };
				\node [ below = 4.500 cm ] at (Classifier) { $\ket{ \psi_{ 3 } }$ };
				\node [ below = 4.500 cm ] at (Measurement) { $\ket{ \psi_{ f } }$ };
			\end{yquant}
		\end{tikzpicture}
		\caption{This figure visualizes the general quantum circuit QCPC$_{ 2 n }$ that implements the BFPQC algorithm for the classification of the functions contained in $F_{ 2 n }$.}
		\label{fig: The BFPQC Quantum Circuit for $F_{ 2 n }$}
	\end{figure}
\end{tcolorbox}

Therefore, the abstract quantum circuit that implements the BFPQC algorithm for the classification of the class of Boolean functions $F_{ 2 n }, n \geq 1$, is outlined in Figure \ref{fig: The BFPQC Quantum Circuit for $F_{ 2 n }$}. To avoid any ambiguity, we explain the notation used in this figure.

\begin{itemize}
	\item	
	$IR$ is the quantum input register that contains $2 n$ qubits and starts its operation at state $\ket{ \mathbf{ 0 } }$.
	\item	
	$OR$ is the single-qubit output register initialized to $\ket{ - }$.
	\item	
	$H$ is the Hadamard transform.
	\item	
	$U_{ f }$ is the unitary transform corresponding to the oracle for the unknown function $f$. The latter is promised to be an element of $F_{ 2 n }$.
	\item	
	$Q_{ 2 }$ is the fundamental building block of $Q_{ 2 n }$, as evidenced by equation \eqref{eq: New Classifiers from Old}.
\end{itemize}

\begin{tcolorbox}
	[
		enhanced,
		breakable,
		center title,
		fonttitle = \bfseries,
		colbacktitle = azure4,
		coltitle = white,
		title = How classification works,
		grow to left by = 0.000 cm,
		grow to right by = 0.000 cm,
		colframe = azure1,
		colback = azure9!75,
		enhanced jigsaw,			
		sharp corners,
		boxrule = 0.500 pt,
	]
	The Boolean functions contained in $F_{ 2 n }$ are enumerated as $f_{ 0 }, f_{ 1 }, \dots, f_{ 2^{ 2 n } -1 }$. Assuming the oracle encodes the function $f_{ i }$ with index $i$, the outcome of the final measurement of the quantum circuit QCPC$_{ 2 n }$ after the action of the classifier $Q_{ 2 }^{ \otimes 2 n }$ will be $\ket{ \mathbf{ i } }$, where $\mathbf{ i }$ is the binary representation of the index $i$, i.e., one of the basis kets of the computational basis. Our algorithm is optimal because it requires just a single query to classify the hidden function.
\end{tcolorbox}

The method we used to devise the BFPQC algorithm is visualized in Figure \ref{fig: BFPQC Methodology Diagram}. We are confident that this methodology is general and fruitful, in the sense that it can be used as a starting point to define additional quantum classification algorithms by establishing different hierarchies of pattern bases and classifiers.

\begin{tcolorbox}
	[
		enhanced,
		breakable,
		center title,
		fonttitle = \bfseries,
		grow to left by = 0.000 cm,
		grow to right by = 0.000 cm,
		colback = WordBlueVeryLight!03,
		enhanced jigsaw,			
		sharp corners,
		toprule = 0.001 pt,
		bottomrule = 0.001 pt,
		leftrule = 0.001 pt,
		rightrule = 0.001 pt,
	]
	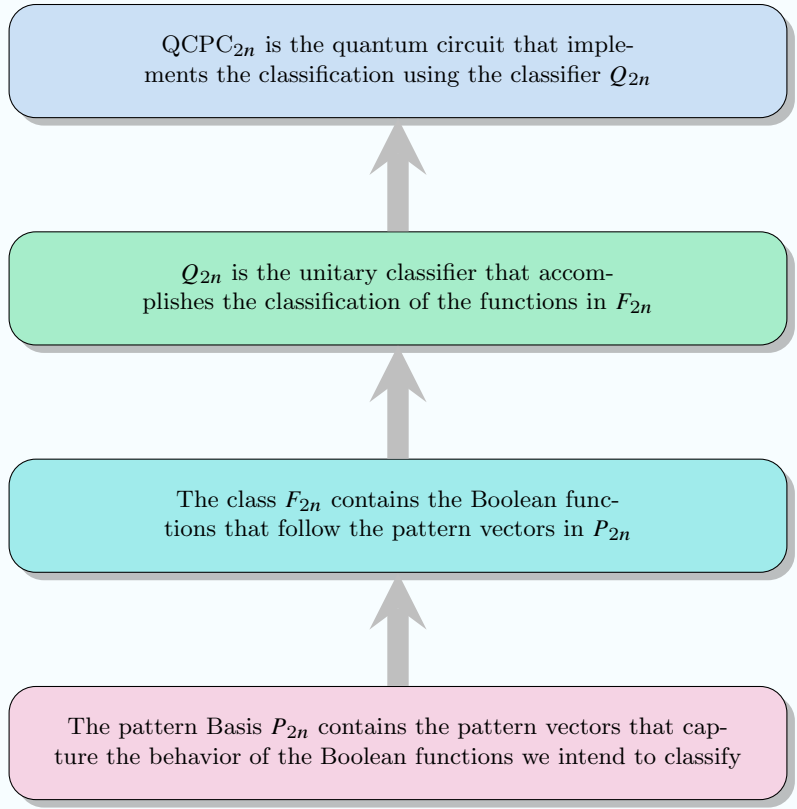
\begin{figure}[H]
		\centering
		\begin{tikzpicture}
			[
			box/.style =
			{
				rectangle,						
				text width = 10.000 cm,			
				minimum height = 1.500 cm,		
				align = center,					
				draw,							
				font = \small,					
				rounded corners = 10.000 pt,	
				drop shadow =
				{
					shadow xshift = 3.000 pt,
					shadow yshift = - 3.000 pt,
					opacity = 0.500,
				},								
			},
			arrow/.style =
			{
				->,								
				thick,							
				>=stealth,						
				line width = 8.000 pt,			
				draw = gray8,					
			}
			]
			\node [ box, fill = purple9 ] (box1) at (0, 0) {
				The pattern Basis $P_{2n}$ contains the pattern vectors that capture the behavior of the Boolean functions we intend to classify
			};
			\node [ box, fill = cyan9, above = 1.500 cm of box1 ] (box2) {
				The class $F_{2n}$ contains the Boolean functions that follow the pattern vectors in $P_{2n}$
			};
			\node [ box, fill = teal9, above = 1.500 cm of box2 ] (box3) {
				$Q_{2n}$ is the unitary classifier that accomplishes the classification of the functions in $F_{2n}$
			};
			\node [ box, fill = azure9, above = 1.500 cm of box3 ] (box4) {
				QCPC$_{2n}$ is the quantum circuit that implements the classification using the classifier $Q_{2n}$
			};
			\draw [ arrow ] (box1) -- (box2);
			\draw [ arrow ] (box2) -- (box3);
			\draw [ arrow ] (box3) -- (box4);
		\end{tikzpicture}
		\caption{This diagram visualizes the main stages of the methodology we employed to create the BFPQC algorithm.}
		\label{fig: BFPQC Methodology Diagram}
	\end{figure}
\end{tcolorbox}

We close this Section by giving a more interesting example targeting functions of $F_{ 4 }$.

\begin{example} {Classifying functions of $F_{ 4 }$} { Classifying $F_{ 4 }$}
Let us assume that Bob has to choose a Boolean function from $F_{ 4 }$, the promised class of functions in this case. Say that Bob chooses $f_{ 3 }$, the behavior of which is given by the pattern vector $1000 \ 1000 \ 1000 \ 0111$, listed in Example \ref{xmp: Pattern Basis $P_{ 4 }$}. Alice makes her move by employing the classifier $Q_{ 4 } = Q_{ 2 }^{ \otimes 2 }$. In this case, the concrete implementation in Qiskit \cite{Qiskit2025} of the general quantum circuit of Figure \ref{fig: The BFPQC Quantum Circuit for $F_{ 2 n }$} takes the form shown in Figure \ref{fig: Phase4___0100___}, where Bob uses the oracle for the function $f_{ 3 }$ and Alice the classifier $Q_{ 4 }$.

\begin{tcolorbox}
	[
		enhanced,
		breakable,
		center title,
		fonttitle = \bfseries,
		grow to left by = 0.500 cm,
		grow to right by = 0.500 cm,
		colframe = cyan4,
		colback = white,			
		enhanced jigsaw,			
		sharp corners,
		toprule = 0.000 pt,
		bottomrule = 0.000 pt,
		leftrule = 0.001 pt,
		rightrule = 0.001 pt,
	]
	\begin{figure}[H]
		\centering
		\includegraphics [ scale = 0.515, trim = {0.850cm 0.000cm 0.000cm 0.750cm}, clip ] { "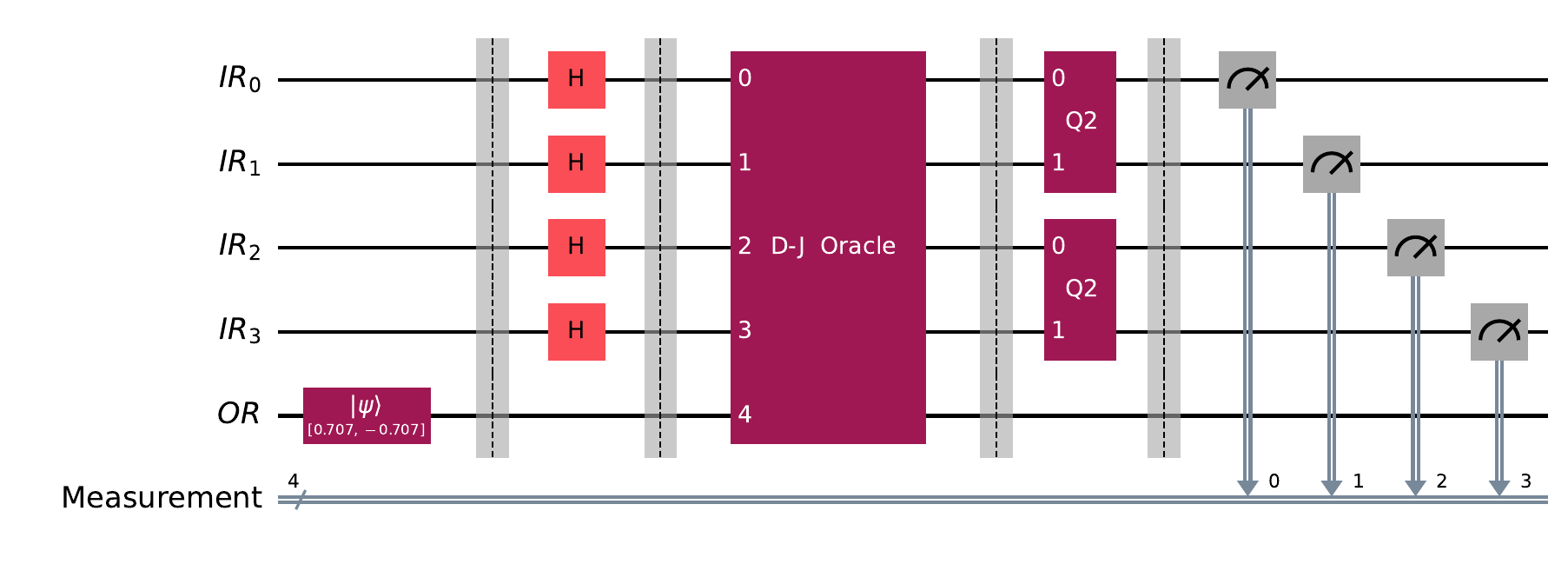" }
		\caption{This figure shows the implementation of the BFPQC algorithm for the classification of the Boolean functions in $F_{ 4 }$, assuming Bob has chosen the oracle for the function $f_{ 3 }$ and Alice has employed the classifier $Q_{ 4 }$.}
		\label{fig: Phase4___1000100010000111___}
	\end{figure}
\end{tcolorbox}

\begin{tcolorbox}
	[
		enhanced,
		breakable,
		center title,
		fonttitle = \bfseries,
		grow to left by = 0.500 cm,
		grow to right by = 1.000 cm,
		colframe = cyan4,
		colback = white,			
		enhanced jigsaw,			
		sharp corners,
		toprule = 0.000 pt,
		bottomrule = 0.000 pt,
		leftrule = 0.001 pt,
		rightrule = 0.001 pt,
	]
	\begin{figure}[H]
		\centering
		\includegraphics [ scale = 0.525, trim = {0.000cm 0.000cm 0.000cm 0.000cm}, clip ] { "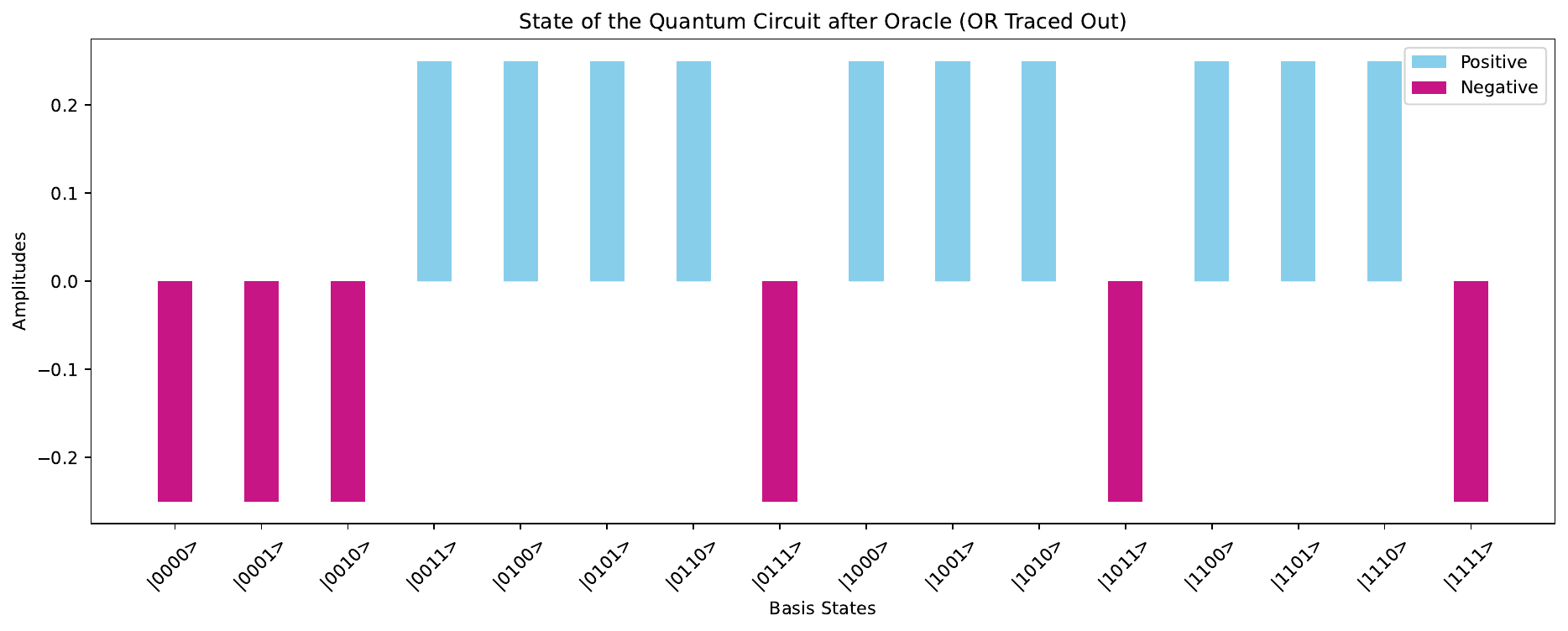" }
		\caption{This is the state of the quantum circuit of Figure \ref{fig: Phase4___1000100010000111___} after the oracle but before the action of $Q_{ 2 }^{ \otimes 2 }$.}
		\label{fig: StateAfterOracle___1000100010000111___}
	\end{figure}
\end{tcolorbox}

After the oracle, and before the action of the classifier, the state of the system is shown in Figure \ref{fig: StateAfterOracle___1000100010000111___}. After the action of the classifier, the state of the system is just $\ket{ \mathbf{ 0011 } }$. Therefore, measuring the quantum circuit depicted in Figure \ref{fig: Phase4___1000100010000111___} will output the bit vector $0011$ with probability $1$ (as corroborated by the measurements contained in Figure \ref{fig: Phase4_Histogram_StatevectorSampler___1000100010000111___}), which is the binary representation of the index of $f_{ 3 }$. Alice surely wind the game, as anticipated.

\begin{tcolorbox}
	[
		enhanced,
		breakable,
		center title,
		fonttitle = \bfseries,
		grow to left by = 0.000 cm,
		grow to right by = 0.000 cm,
		colframe = cyan4,
		colback = white,			
		enhanced jigsaw,			
		sharp corners,
		toprule = 0.000 pt,
		bottomrule = 0.000 pt,
		leftrule = 0.001 pt,
		rightrule = 0.001 pt,
	]
	\begin{figure}[H]
		\centering
		\includegraphics [ scale = 0.600, trim = {0.000cm 0.000cm 0.000cm 0.000cm}, clip ] { "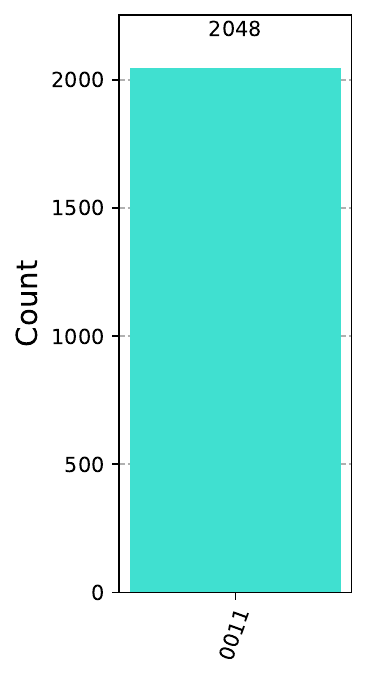" }
		\caption{This is the measurement outcome of the quantum circuit of Figure \ref{fig: Phase4___1000100010000111___} for $2048$ runs.}
		\label{fig: Phase4_Histogram_StatevectorSampler___1000100010000111___}
	\end{figure}
\end{tcolorbox}

\end{example}

\section{Discussion and conclusions} \label{sec: Discussion and Conclusions}

The literature is full of advanced studies that extend and generalize the Deutsch-Jozsa algorithm and investigate balanced Boolean functions. Nevertheless, to the best of our knowledge, there has not been any prior research focusing on imbalanced Boolean functions, which are characterized by an unequal number of elements in their domain that take the values $0$ and $1$. This article presents a new quantum algorithm aimed at categorizing a particular hierarchy of imbalanced Boolean function classes.

For each positive integer $n \geq 1$, this hierarchy encompasses a class $F_{n}$ of $n$-ary Boolean functions, which are delineated based on their behavioral traits. A distinguishing characteristic of all functions within the same class is their common imbalance ratio. Our algorithm enables classification in a clear way, as the final measurement identifies the unknown function with a probability of $1$. It is crucial to emphasize that the proposed algorithm is an optimal oracular algorithm, capable of categorizing the specified functions with a single query to the oracle.

Let us note that, as previously explained, within the quantum context $f_{ i }$ and its negation $\overline{ f_{ i } }$ are indistinguishable because they lead to the same state. To distinguish between them, if need be, will require a second query to the oracle for a single specific input value $\mathbf{ x }$ because the outcome will conclusively differentiate $f_{ i }$ from $\overline{ f_{ i } }$.

In closing, we emphasize that, in addition to a concrete algorithm, at the beginning of Section \ref{sec: The Basic Concepts Behind the BFPQC Algorithm} we offer a comprehensive description of the methodology utilized in the creation of this algorithm. This is done with the hope that it will prove both general and advantageous, as it can be easily modified and expanded to address other categories of imbalanced Boolean functions that exhibit diverse behavioral patterns.

\bibliographystyle{ieeetr}
\bibliography{BFPQC}

\end{document}